 \documentclass[%
  reprint,
 superscriptaddress,
  amsmath,amssymb,
  aps,
 ]{revtex4-1}
\usepackage{epsfig}
\usepackage{times}
\usepackage{amsmath}
\usepackage{amsfonts}
\usepackage{amssymb}
\usepackage[usenames]{color}
\usepackage{graphicx}
\usepackage{graphicx}
\usepackage[dvipdfmx]{}
\usepackage{bm}
\usepackage{xcolor}
\usepackage{braket}
\usepackage{here}
\begin{document}

\title{Quantum metrology with generalized cat states 
}

\author{Mamiko Tatsuta}
\email[email address:]{mamiko@as.c.u-tokyo.ac.jp}
\affiliation{Komaba Institute for Science, The University of Tokyo, 3-8-1 Komaba, Meguro, Tokyo 153-8902, Japan}
\affiliation{Department of Basic Science, The University of Tokyo, 3-8-1 Komaba, Meguro, Tokyo 153-8902, Japan}
\author{Yuichiro Matsuzaki}
\email[Current address: Nanoelectronics Research Institute, National Institute of Advanced Industrial Science and Technology (AIST), 1-1-1
Umezono, Tsukuba, Ibaraki 305-8568, Japan]{}
\affiliation{NTT Basic Research Laboratories, NTT Corporation, 3-1 Morinosato-Wakamiya, Atsugi, Kanagawa, 243-0198, Japan}
\author{Akira Shimizu}%
\email[email address:]{shmz@as.c.u-tokyo.ac.jp}
\affiliation{Komaba Institute for Science, The University of Tokyo, 3-8-1 Komaba, Meguro, Tokyo 153-8902, Japan}
\affiliation{Department of Basic Science, The University of Tokyo, 3-8-1 Komaba, Meguro, Tokyo 153-8902, Japan}




\date{\today}

\begin{abstract}
 We show a general relationship between a superposition
 of macroscopically distinct states and sensitivity in quantum
 metrology. 
 Generalized cat states 
are
 defined by
 using 
an index 
which
 extracts the coherence between macroscopically
 distinct states, and a wide variety of states, including a classical
 mixture of an exponentially large number of states, has been identified
 as the generalized cat state with this criterion.
 We find that if we use the generalized cat states for magnetic field
 sensing without noise, the sensitivity achieves the Heisenberg scaling.
 More importantly, 
we even show that 
sensitivity of generalized cat states achieves the ultimate scaling sensitivity beyond  the standard quantum limit 
 under the effect of dephasing.
As an example, we investigate the sensitivity of a generalized cat state 
that is attainable through a single global manipulation on 
a thermal equilibrium state 
 and find an improvement of a few orders of magnitude from the previous sensors.
Clarifying a wide class that includes such a peculiar state as metrologically useful, our results significantly broaden the potential of quantum metrology.
\end{abstract}

\pacs{Valid PACS appear here}
\maketitle


\section{Introduction}
High-precision metrology is important in both fundamental and applicational senses \cite{giovannetti2004quantum,giovannetti2011advances,taylor2016quantum,degen2017quantum}. 
In particular,
magnetic field sensing has been attracting much attention
\cite{wineland1992spin,wineland1994squeezed,toth2014quantum}
due to the potential applications in various fields 
 from  the determination of the
structure of chemical compounds to imaging of living cells \cite{le2013optical}. 
Numerous efforts have
been made to increase the sensitivity of the magnetic field sensors \cite{paris2009quantum,
chin2012quantum,chaves2013noisy,Jones1166,
huelga1997improvement,kuzmich1998atomic,fleischhauer2000quantum,geremia2003quantum,leibfried2004toward,auzinsh2004can,dunningham2006using,matsuzaki2011magnetic,demkowicz2012elusive,bohnet2014reduced,tanaka2015proposed,
dooley2016hybrid,davis2016approaching,matsuzaki2018quantum}, and
various types of magnetometers have been studied  \cite{huber2008gradiometric,ramsden2011hall,poggio2010force}.
 A qubit-based sensing \cite{happer1973spin,allred2002high,dang2010ultrahigh,bal2012ultrasensitive,toida2017electron,acosta2009diamonds,balasubramanian2009ultralong,nvcenter,ishikawa2012optical} 
 is an attractive
approach where quantum properties are exploited to enhance the
sensitivity. 
By using superpositions of states,  the standard Ramsey-type
measurement without feedback
can be implemented to measure the magnetic field, where the
magnetic field information is encoded in the relative phase  between the
states in accordance with the magnetic field strength. 
 If we use $N$  qubits in separable states, it is known that 
the uncertainty
(that is, the inverse of the sensitivity) scales as $\Theta(N^{-1/2})$, which
is called the standard quantum limit (SQL) \footnote{As done in the field of quantum metrology, we focus on the scaling, neglecting the constant factor.}. On the other hand, quantum physics
allows one to beat the SQL.  
The ultimate scalings are known to be $\Theta(N^{-1})$, i.e., the Heisenberg scaling, in the absence of noise and $\Theta(N^{-3/4})$ in the presence of realistic decoherence \cite{Jones1166,chin2012quantum,matsuzaki2011magnetic,tanaka2015proposed,dooley2016hybrid,matsuzaki2018quantum,palma1996quantum,smirne2016ultimate,macieszczak2015zeno}.

In the standard Ramsey-type measurement protocol,
the ultimate scalings seem 
 to be attainable by using the quantum superposition. However, a general relationship between a quantum superposition and sensitivity is not yet known. Therefore, it is essential to clarify what type
of superposition 
gives higher sensitivity in metrology than classical sensors.

Superpositions of macroscopically distinct states, i.e., 
``cat'' states, have attracted many researchers due to the
fundamental interest since its introduction 
by 
Schr\"odinger \cite{schroedinger}. Although a cat state
contains a superposition, not all types of superpositions can be considered
as the cat state.   
The Greenberger-Horne-Zeilinger (GHZ)
 \cite{greenberger1990bell,monz201114,dicarlo2010preparation} state is one of the typical cat states.
Since this 
 cat state is useful in quantum metrology, we may expect other cat states to be useful as well.
 However, there
was no unified criteria to judge if a given state contains such 
macroscopically distinct states \cite{frowis2018macroscopic}, preventing the further understanding of the relation between cat states and sensors.
Among many possible measures, we especially focus on  the
index $q$ \cite{q}. 
Importantly, 
 $q$ is defined for both pure
 and mixed states, and is measurable in experiments by  measuring a certain set of local observables.

\begin{figure}
 \centering
 \includegraphics[keepaspectratio, scale=0.65]
      {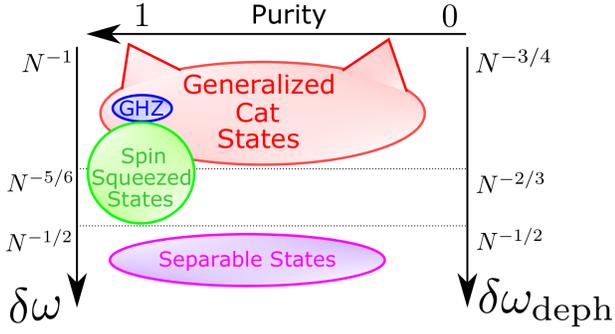}
 \caption{
The relationship between the purity and the scaling of the uncertainty for given quantum states when  we use the quantum states for the Ramsey-type quantum sensing. 
The ultimate scaling of the uncertainty without [with] dephasing is  $\delta \omega =\Theta (N^{-1})$ [$\delta\omega_{\rm deph}=\Theta(N^{-3/4})$].
Only special pure entangled states such as GHZ states are known to achieve such a scaling.
The GHZ state is a pure state, and the uncertainty  scales as $\Theta(N^{-1})$ in the absence of dephasing and $\Theta(N^{-3/4})$ in the presence of dephasing. One-axis and two-axis spin squeezed states \cite{kitagawa1993squeezed} are pure states beating the SQL. Separable states, whether pure or mixed, do not beat the SQL. 
In this paper, we show that all the generalized cat states achieve the ultimate scalings, even if it is a classical mixture of exponentially large number of  states.}
\label{statekando}
\end{figure}

In this paper, we prove that generalized cat states, i.e., the superposition of macroscopically distinct states characterized by the index $q$, are all capable of achieving 
the ultimate scalings.
We give the upper bound of the uncertainty when $q=2$ states are used as a sensor state.
First, we show the Heisenberg  scaling in the absence of noise.
Second, we analyze the case with a realistic decoherence. 
We prove that the SQL is still beaten; 
the generalized cat states  achieve the ultimate scaling uncertainty $\Theta(N^{-3/4})$. 
Third, we present a nontrivial example and numerically show its  advantage. 
Since there are states with low purity  among the generalized cat states (Fig.~\ref{statekando}), 
wide varieties of states have the potential to
 achieve the ultimate scalings.

\section{Generalized cat states}
To begin  with, we introduce a
concept of a
{\it generalized cat state}, which is discussed in detail in the appendix  of \cite{tatsuta2018conversion}.
We refer to the index $q$ \cite{q,frowis2012measures,morimae2010superposition,toth2014quantum,jeong2015characterizations,frowis2015linking,PhysRevB.93.195127}, which is a real number satisfying $1 \leq q \leq 2$.
It is defined as
\begin{align}
\max\{N,\max_{\hat A, \hat \eta}\mathrm{Tr}\left(\hat\rho\big[\hat A,\big[\hat A, \hat \eta \big]\big]\right)\}
=\Theta (N^q)\label{qdef},
\end{align}
where $\hat A=\sum_{l=1}^N\hat a(l) $ is an additive observable and $\hat \eta$ is a projection operator.
Since the states with $q=2$ have the interesting features that we would like to focus on in this paper,
 we 
 simplify the definition for this case as follows.
A quantum state $\hat \rho $ has $q=2$
if there exist an additive observable 
$\hat A$
 and a projection operator $\hat \eta$ such that 
 \begin{align}
 \mathrm{Tr}(\hat \rho[\hat A,[\hat A,\hat \eta]])=\Theta(N^2).\label{def:q2}
 \end{align}
We call a state  with $q=2$ a generalized cat state.
 By contrast, e.g., 
separable states have $q=1$. 

We can understand the physical meaning of $q$
by 
 expressing  the left hand side of Eq.~(\ref{def:q2}) as follows:
$\mathrm{Tr}(\hat \rho[\hat A,[\hat A,\hat \eta]])=\sum_{A,\nu,A',\nu'}(A-A')^2\braket{A,\nu|\hat \rho|A',\nu'}\braket{A',\nu'|\hat \eta|A,\nu}$, where $\ket{A,\nu}$ denotes an eigenvector of $\hat A$ with eigenvalue $A$, and
$\nu$ denotes the degeneracy. This shows that, 
if 
$\hat \rho$  has $q=2$,
there exist terms such that $\braket{A,\nu|\hat \rho|A',\nu'}\braket{A',\nu'|\hat\eta|A,\nu} \neq 0$ for $|A-A'|=\Theta(N)$.
For $N \gg 1$,
the term $\braket{A',\nu'|\hat\rho|A,\nu}$ with $|A-A'|=\Theta(N)$ 
corresponds to a quantum
coherence between 
states that 
are  distinguishable even 
on  a macroscopic 
 scale. 
Therefore,
the state with $q=2$
can be considered to contain a superposition of macroscopically distinct states. 


 For pure states, $q=2$ guarantees the existence of an additive observable such that ${\mathrm{Tr}}[ \hat \rho (\Delta \hat{A})^2 ]=\Theta (N^2)$. As suggested from other measures of macroscopic quantum states \cite{park2016quantum,frowis2012measures}, such a large fluctuation is available only when $\hat \rho$ has a superposition of macroscopically distinct states (for details, see the appendix of \cite{speedup}). 
As an example, let us consider a state $\ket{ \psi }
:=
(
\ket{\downarrow}^{\otimes N}
+
\ket{\uparrow}\ket{\downarrow}^{\otimes N-1}
+
\ket{\uparrow}^{\otimes 2}\ket{\downarrow}^{\otimes N-2}
+
\cdots
+
\ket{\uparrow}^{\otimes N}
)/ \sqrt{N+1}$. Since this state is much more complicated than the
well-known
GHZ state, it may be difficult to 
intuitively 
judge whether this is a cat
state, but we can actually show that this state has $q=2$ 
by taking $\hat A=\hat M_z$ and $\hat\eta=\ket{\psi}\bra{\psi}$.
Pure states with $q=2$ are known to have 
several ``cat like''
properties, such as fragility against decoherence and instability against
local measurements \cite{shmzmiyadera2002}.

For mixed states, $q$ correctly identifies states that contain pure cat states with a significant ratio in the following sense
 (see, e.g., the appendix of \cite{tatsuta2018conversion}).
Without losing generality, we can
perform
a pure state decomposition of a mixed state with $q=2$  as
$\hat{\rho} = \sum_{j=1}^N \lambda_j \ket{\psi_j}\bra{\psi_j}$,
where $\ket{\psi_j}\bra{\psi_j}$ has $q=2$ ($q<2$) for $j=1,2,\cdots ,m$ ($j=m+1,m+2,\cdots ,N$) for $0<m<N$.
In this case,  we can show $\sum_{j=1}^{m} \lambda _{j}=\Theta(N^0)$, and this intuitively means that a mixed state with $q=2$ contains a significant (or nonvanishing) amount of pure states with $q=2$. 
For example, $\hat \rho_{\rm ex} = w  \ket{\psi}\bra{\psi} + (1-w) \hat \rho_{\rm sep}$ has $q=2$ for $N$-independent $w>0$, where  $\hat \rho_{\rm sep}$ is an arbitrary separable state.


\section{Definition of sensitivity}
Since we will later discuss the relationship between the
generalized cat states and quantum sensing, we
 review the concept of quantum metrology. 
Here we discuss
the case of  a spin 
 system to exemplify in the
context of magnetometry, although our results are, in principle, applicable to any physical systems, e.g.,  
interferometry in optical systems \cite{giovannetti2011advances}.
 Suppose that a sensor consists of
$N$ free spins that interact with 
 a 
 magnetic field with a
Hamiltonian 
$\hat H_0(\omega)=\omega \hat A$, where $\omega $ denotes 
the Zeeman frequency shift of the spins
and $\hat A$ is the sum of local spin operators [hence $\|\hat A\|=\Theta(N)$].
 We assume that the  frequency has a linear scaling
with respect to the magnetic field $B$ (such as $\omega \propto B$).
 Also, we decompose magnetic field $B$ into  the
 ``applied field'' $B_0$ (corresponding Zeeman shift $\omega _0$) and the
 ``target field'' $B'$ (corresponding Zeeman shift $\omega '$); $\omega=\omega_0+\omega'$. Here, we
 assume that we know the amplitude of the applied magnetic field $B_0$
 while the target small magnetic field $B'$ is unknown.
For metrological interest, we consider   $\omega'\rightarrow 0$ throughout this paper.
Also, to include the effect of the dephasing, we 
add the noise
effect 
to
 the total Hamiltonian as
$\hat H=\hat H_0(\omega)+\hat H_{\rm{int}}$, where $\hat H_{\rm{int}}$
denotes the interaction  with the environment.

The following is 
the standard Ramsey-type
protocol to detect the
magnetic field by using spins.
First, prepare the spins in the state $\hat \rho$.
Second, let $\hat \rho$ evolve under the Hamiltonian 
$\hat {H}$
for an interaction time $t_{\rm int}$ to become $\hat \rho(t_{\rm int})$.
Third, read out 
the state via a measurement described by
a projection operator $\hat {\mathcal{P}}$
. 
Fourth,  repeat these three steps
within a given total measurement time $T$.
We assume that state preparation  and projection 
can be performed 
 in a short time interval much smaller than $ t_{\rm int}$.
In this case, the number of the repetition is
approximated 
to be
 $T/t_{\rm int}$, and therefore the uncertainty  $\delta \omega$
of the estimation
of our protocol is described as 
\begin{align}
\delta \omega = 
\frac{\sqrt{P(1-P)}}{\left|\frac{dP}{d\omega}\right|}\frac{1}{\sqrt{T/t_{\rm int}}},
\label{eq:delta_omega}
\end{align}
where $P=\mathrm{Tr}(
\hat \rho(t_{\rm int}) 
\hat {\mathcal{P}}
)$
denotes the probability that the projection described by 
$\hat {\mathcal{P}}$
occurs
at the readout process.

\section{Heisenberg scalingt in the ideal environment}
Here, we show that we can achieve the Heisenberg scaling, i.e., $\Theta(N^{-1})$ uncertainty,
by using a state with $q=2$ as a sensor of the target  field if
decoherence is negligible.


Suppose that we have a generalized cat state $\hat{\rho }$ satisfying
Eq.~(\ref{def:q2})
for
an additive observable $\hat A$ and
a projection operator $\hat \eta$.
If the target field couples with the spins via $\hat A$ as 
$\hat H_0(\omega)=\omega \hat{A}$, which induces 
an energy change,
we can use the state with $q=2$ to sensitively estimate the value of
$\omega $. 
By setting the projection operator for the readout as
$\hat{\mathcal{P}}=\hat{\eta }$, we can use the standard sensing protocol described in the previous paragraph.
We  find that for a  certain positive constant $p_1$, there
exist $\Omega_1=\Theta(N^0)$ and  $N_1 >0$ such that
\begin{align}
\delta \omega &\leq 
\left(p_1 p_2
^2Nt_{\rm int}\right)^{-1}\left(\sqrt{T/t_{\rm int}}\right)^{-1}
\end{align}
is satisfied for $p_2:=\omega t_{\rm int} N =\Theta(N^0)\leq \Omega_1$ and $N\geq N_1$. 
This is because the numerator  of Eq.~(\ref{eq:delta_omega}) 
satisfies $\sqrt{P(1-P)}=\Theta(N^0)$ for $\omega t_{\rm int} N=\Theta(N^0)$, 
 whereas $|dP/d\omega|$ in
the denominator 
has a lower bound;
\begin{align}
\left|\frac{dP}{d\omega}\right|\geq
\left| \left|\omega t_{\rm int}^2\mathrm{Tr}(\hat\rho[\hat A,[\hat A,\hat\eta]])\right|
-\left|it_{\rm int}\mathrm{Tr}(\hat\rho [\hat A,\hat \eta])\right|
\right|\nonumber\\
-2t_{\rm int}\|\hat A\|(e^{2\omega t_{\rm int}\|\hat A\|}-1-2\omega t_{\rm int}\|\hat A\|).\label{bound}
\end{align}
Since we assume Eq.~(\ref{def:q2}), the term $u:=\left|\omega t_{\rm int}^2\mathrm{Tr}(\hat\rho[\hat A,[\hat A,\hat\eta]])\right|=p_2
\Theta(t_{\rm int} N)$, whereas
the term $v:=\left|it_{\rm int}\mathrm{Tr}(\hat\rho [\hat A,\hat \eta])\right| \leq \Theta(t_{\rm int} N)$.
Therefore,  we obtain
 $\left|u-v
\right|=p_2
\Theta(t_{\rm int} N)$ by tuning $p_2
=\Theta(N^0)<1$ correctly.
The remaining term in Eq.~(\ref{bound}) is $-\Theta(t_{\rm int} N)p_2
^2$, which can be made much smaller than  $|u-v|$  by taking  $p_2
\ll 1$. More precisely,  
 we find that there exists a positive constant $\Omega_1 \ll 1$ such that $\forall p_2
=\Theta(N^0) \leq \Omega_1$ satisfies $|dP/d\omega|\geq p_1p_2
^2t_{\rm int} N$ for a certain positive constant $p_1$.
If we  tune $\omega_0$  in such a way that $\omega=\omega_0+\omega'$ scales as $\omega =\Theta(N^{-1})$,
and choose the interaction time $t_{\rm int}=\Theta(N^0)$ as to realize the condition of $\omega t_{\rm int} N=\Theta(N^0)$,
then we have
$\delta \omega \leq 1/\Theta(N)$, 
achieving 
the Heisenberg  scaling.

\section{Ultimate scaling in the presence of decoherence}
In reality, dephasing is one of the major challenges to be overcome for  beating the SQL. 
For example, the GHZ state acquires
 the  information of the target field 
as a relative phase $\exp(i\omega' t N)$ on the off-diagonal terms of the density matrix.
However, the dephasing induces a rapid decay of the amplitude of such off-diagonal terms, making it nontrivial  whether or not the quantum sensor really has an advantage. 

Upon discussing the dephasing, we must take into account 
the correlation time $\tau_c$ of the environment.
Historically, 
 the Markovian dephasing was considered 
for evaluating the performance of the quantum 
sensor \cite{huelga1997improvement,escher2011general,demkowicz2012elusive,kolodynski2013efficient}.
This implies that 
$\tau_c$ was assumed to be much smaller than any other time 
scales such as the coherence time $T_2^*$ and $t_{\rm int}$.
Then, 
if we reasonably assume the independent dephasing, the 
decay of the  off-diagonal terms behaves 
as $\exp(-tN/T_2^*)$, which is not slower than the phase accumulation
$\exp(i\omega' t N)$.
In this case, it was concluded that beating the SQL is impossible even with the optimal interaction time [which is
 $t_{\rm int} =\Theta(1/N)$
].

However, in  most of the solid-state  qubits,
$\tau_c \gg T_2^*$
in contradiction to the Markovian dephasing.
By taking this point into account, 
Refs.~\cite{chin2012quantum,palma1996quantum,Jones1166,matsuzaki2011magnetic,tanaka2015proposed,matsuzaki2018quantum} recently found that 
$t_{\rm int}$ should be taken in the so-called Zeno regime, 
i.e., $t_{\rm int} \ll \tau_c$, 
where the non-Markovian effect plays a crucial role.
The decay of the off-diagonal terms in this regime 
 behaves as $\exp(-(t/T_2^*)^2N)$, which  is much slower than the decay in Markovian dephasing.
With the optimal interaction time $t_{\rm int}  \sim T_2^*/\sqrt N$, 
 it was proven that the GHZ state and spin squeezed states 
can beat the SQL, achieving the ultimate scaling $\delta\omega\propto N^{-3/4}$  \cite{Jones1166,chin2012quantum,matsuzaki2011magnetic,tanaka2015proposed,dooley2016hybrid,matsuzaki2018quantum,palma1996quantum}. However, these investigations were limited to some 
specific 
 states,
leaving an open question of whether or not there are any other metrologically 
useful 
superpositions. Moreover, although 
most of the previous research
 assumed that  pure states can be prepared, 
quantum
 states for sensing may be mixed in experiments. 
So, for understanding the full potential of quantum metrology, 
it is crucial to explore the sensitivities of sensing using other, 
nontrivial and non-ideal, states.

Here, we discuss the performance of the generalized cat
states
satisfying Eq.~(\ref{def:q2}) 
 as a magnetic field sensor under the effect of 
 dephasing with 
$\tau_c$ longer than $t_{\rm int}$. 
We model the dephasing by 
adding Hamiltonian $\hat H_0(\omega)$ 
the following
interaction 
with the environment \cite{palma1996quantum,hornberger2009introduction}:
$\hat H_{\rm{int}}=\sum_{l=1}^N \lambda f_l(t)\hat a(l)$,
where $\lambda $ denotes the amplitude of the noise and $f_l(t)$
$(l=1,2,\cdots ,N)$ denotes
a random classical variable at the site $l$.
We assume   $f_l(t)$ satisfies
$\overline{f_l(t)}= 0$ 
and
$\overline{f_l(t)f_{l'}(t')} = \exp(-|t-t'|/\tau_c)\delta_{l,l'}$,
where the overline denotes the
ensemble average. 
Taking $t_{\rm int} \ll \tau_c$,
 we can approximate $\exp(-|t-t'|/\tau_c)\simeq 1$ because $|t-t'|\leq t_{\rm int}$.
When there is such a dephasing, the state after the time evolution is a
classical mixture of $\exp(-i\omega \hat A
t_{\rm int})\hat\rho\exp(i\omega \hat A t_{\rm int})$ [with a weight of $(\frac{1+\exp(-2\lambda^2t_{\rm int}^2)}{2})^N$]
 and other states. The former state corresponds to the
 generalized cat state that has evolved 
in
 the magnetic field without dephasing.
 Although we have shown that the former
state can achieve the Heisenberg scaling, the latter state has a
complicated form, and so the calculation of the sensitivity of the
latter state is not
straightforward. 
Fortunately, by tuning 
$p_2(=\omega t_{\rm int}
N=\Theta(N^0)\ll 1$) and $t_{\rm int}$, the former contribution can be set to be larger
than the latter contribution, and the uncertainty can be bounded as follows: 
\begin{align}
\delta \omega_{\rm deph}\sqrt T
& \leq
(N\sqrt{t_{\rm int}})^{-1}
\Big[ p_1 p_2^2 
\Big(\frac{1+e^{-2\lambda^2t_{\rm int}^2}}{2}\Big)^N
\nonumber\\
&-2e^{2\omega t_{\rm int} \|\hat A\|
}\frac{\|\hat A\|}{N} \Big(1-\Big(\frac{1+e^{-2\lambda^2t_{\rm int}^2}}{2}\Big)^N \Big) \Big]^{-1}.
\label{withdephasing} 
\end{align}
By taking $t_{\rm int}\propto p_2^2
/\sqrt{N}$, 
we obtain
 $\delta \omega_{\rm deph} \sqrt{T}\leq \Theta(N^{-3/4})$, and this
achieves the ultimate scaling beyond the SQL.
We can see the optimality of 
this scaling  as
follows. As we increase $t_{\rm int}$,  the term
$(N\sqrt{t_{\rm int}})^{-1}$ on the right hand side of (\ref{withdephasing})  becomes
smaller,  which contributes to achieve a better sensitivity. 
However, since we need to have a finite weight of $\exp(-i\omega \hat A
t_{\rm int})\hat\rho\exp(i\omega \hat A t_{\rm int})$, its
weight
$\left(\frac{1+e^{-2\lambda^2t_{\rm int}^2}}{2}\right)^N$ 
should be nonvanishing in the limit of large $N$, 
hence scaling of $t_{\rm int}$ should be  $\Theta(1/\sqrt N)$ at most.
Also, we should tune $t_{\rm int}\propto p_2^2$ so the right hand side of Eq.~(\ref{withdephasing}) is positive.
Thus we find  $t_{\rm int}\propto p_2
^2/\sqrt{N}$ is optimal.
Then the scaling of the sensitivity is enhanced $N^{1/4}$ times more than
that of the SQL,
 agreeing with Refs.~ \cite{chin2012quantum, Jones1166, matsuzaki2011magnetic,tanaka2015proposed}, in which  the GHZ beats the SQL by a factor of $N^{1/4}$ with $t_{\rm int}\propto 1/\sqrt{N}$.
Other works also showed 
that
 this scaling is the best in the presence of dephasing \cite{smirne2016ultimate,macieszczak2015zeno}.
Therefore, we have proven that the generalized cat
states can achieve the sensitivity with $\delta \omega_{\rm deph} =\Theta(N^{-3/4})$ that is considered as the ultimate
scaling under the effect of dephasing.

\section{Example}
 We  now discuss a possible application of our results to realize a sensitive magnetic field sensor by using a current technology.
 Recently, it was found that a single measurement 
of the total magnetization $\hat M_z$
converts
a certain thermal
equilibrium state 
into
 a generalized cat state
\cite{tatsuta2018conversion}. The conversion procedure consists of two steps: (1)
apply a magnetic field along a specific direction (that we call the $x$ axis) and let the system equilibrate,
(2) perform a projective measurement $\hat \eta_z$ on $\hat M_z=M$ subspace, where the
$z$ axis is defined as an orthogonal direction to the applied magnetic field.  Then the postmeasurement state has $q=2$ for $M\neq \pm N+o(N)$.
Obviously, for finite temperature, the premeasurement state is a mixture of $\exp[\Theta(N)]$
states because it is a Gibbs state, and the projection measurement is a
projection onto a subspace with dimension of $\exp[\Theta(N)]$. 
This means that the postmeasurement state is a
 mixture of an exponentially large number of states.
Since this state can be prepared from 
a thermal
 equilibrium
 state, this protocol 
has a potential of 
generating
  metrologically useful
 states easily at moderate temperature. 
 Below
we 
discuss the sensitivity when
 we use this state for the sensing $\hat M_x$ with the readout projection $\hat \eta_z$. 

Let us consider phosphorus donor electron spins with the density of $\sim 10^{15}$cm$^{-3}$
in a
$^{28}$Si substrate with a size of $32 \mu$ m $\times 32 \mu$ m$\times
1\mu$m.
Then there are approximately $N=10^6$ electron spins in the substrate.
We assume the applied magnetic field is $10$mT
and the temperature is $10$mK,
where the thermal energy ($k_BT/2\pi \simeq 208$ MHz)
 is comparable with the Zeeman splitting ($g\mu _bB/2\pi \simeq 280$
 MHz) so that the spin is not fully polarized.
 Via a projective
measurement of the total magnetization (that can be implemented by a
superconducting circuit, for example), we can prepare the generalized
cat state with $q=2$. With the coherence time 
  of one electron in this system being around $10$s \cite{eec6e647c2824bbab76a44bb03dd0eeb},
we numerically optimize the interaction time and find that the uncertainty takes its minimum $\delta \omega_{\rm deph} \sqrt{T} =5.2\times 10^{-5}/\sqrt{\rm{Hz}}$ at $t_{\rm int} = 5.4$ms, which corresponds to $\delta B\sqrt{T}=0.30$fT/$\sqrt{\rm{Hz}}$. The optimal interaction time $t_{\rm int} = 5.4$ms is 
consistent with our theoretical prediction that 
 $t_{\rm int}$ should be comparable with the coherence time  divided by $\sqrt N$.
As a comparison, we consider using a 
thermal
equilibrium state 
 in the same conditions as above without 
converting it into the generalized cat state, and we obtain
 $\delta\omega_{\rm deph}\sqrt{T}=9.8\times 10^{-4}/\sqrt{\rm{Hz}}$.  This shows that the use of
 the generalized cat states provides us with 20 times better sensitivity
 than the classical states with this system, which demonstrates the
 practical advantage of the metrology using the generalized cat states.

Let us  compare our results with known theoretical results.
If a fully polarized separable state with the same electron spins is
used, $\delta \omega_{\rm deph} \sqrt T= 8.1\times 10^{-4}/\sqrt{\rm{Hz}}$ is  estimated \cite{tanaka2015proposed}. Also, by squeezing the fully polarized spin state via
nonlinear interactions, it is, in principle, possible to achieve a
sensitivity of $\delta \omega_{\rm deph}\sqrt T = 7.1\times 10^{-5}/\sqrt{\rm{Hz}}$ \cite{tanaka2015proposed},
and this sensitivity is comparable to our results.
However,
these proposals can be implemented only if a perfect initialization of
the electron spins is available, which could be difficult due to the
small Zeeman energy of the electron spins. On the other hand, the sensor state we discuss in this section, 
i.e.,
a generalized cat state in the Si substrate at finite temperature, is initially a thermal equilibrium spin state 
with the polarization ratio around $0.6$, which is more feasible to prepare.
This clearly shows the advantage to use our
generalized cat states.
According to the size of the substrate, the spatial resolution of the sensor is $\sim 10^{-5}$m.
Experimentally achieved
sensitivities with
similar spatial resolution are as follows. A superconducting flux qubit,
a  superconducting quantum interference device (SQUID), and an ensemble
of NV centers showed sensitivities of $3.3$pT$/\sqrt{\rm{Hz}}$ with $5\mu$m resolution \cite{bal2012ultrasensitive}, $1.4$pT$/\sqrt{\rm{Hz}}$ with $100\mu$m resolution
\cite{baudenbacher2003monolithic}, and $150$fT$/\sqrt{\rm{Hz}}$ with $100\mu$m resolution \cite{acosta2009diamonds,le2013optical}, respectively.
Therefore, we can conclude that our proposed sensor has a sensitivity of
at least a few orders of magnitude better than those of the previous
 sensors. 

\section{Discussion}
Although we have mainly discussed the scaling of  $\delta\omega_{\rm deph}$, the quantitative upper bound of $\delta\omega_{\rm deph}$ can be obtained by evaluating the formula (\ref{upperbound}) in the Appendix.

Let us discuss the relation with the quantum Fisher information (QFI).
For a given state, the QFI
gives the {\em lower} bound of $\delta\omega$ as
$\delta\omega \geq 1/\sqrt {\rm {QFI}}$, i.e.,  the Cramer-Rao inequality \cite{paris2009quantum}. 
The equality is satisfied by {\em some} optimal positive-operator valued measure (POVM) operators. However, such operators are generally {\em unknown} for mixed states, and so is the physical measurement process to construct the POVM.
 Hence, practically, the QFI gives $\delta\omega > 1/\sqrt {\rm {QFI}}$,
which does not ensure the ultimate scaling even when 
QFI$=\Theta(N^2)$.
In comparison, we have derived 
the {\em upper} bound of $\delta\omega$ as
$\delta\omega \leq \Theta(N^{-1})$ or $\Theta(N^{-3/4})$ 
for states with $q=2$
assuming a {\em known} measurement: 
the simple Ramsey-type protocol and 
 reading out with the projection $\hat \eta$.  
That is, the way of achieving the ultimate scaling sensitivity is explicitly given.

In addition, 
the dynamical aspects in the presence of noise are not clear enough for the QFI
because in the Cramer-Rao inequality the QFI is of
the state {\em after} the noisy time evolution,
which is not directly related to the QFI of the 
initial state. 
By contrast, 
we have obtained 
the upper bound of $\delta\omega$ 
in terms of $q$ of the {\em initial} cat state,
which is actually prepared in experiments.
Such a practical bound is derived because
$q$ is directly connected to the equation of motion.

\section{Conclusion}
Summing up, we have shown 
that the sensitivity of generalized cat states composed of $N$ spins
 can achieve the Heisenberg
 scaling $\delta \omega=\Theta(N^{-1})$
 if they are used to measure  a magnetic field without
 dephasing.
 Moreover, even in the presence of independent dephasing, we obtained 
 the ultimate scaling $\delta \omega_{\rm deph}=\Theta(N^{-3/4})$ 
beyond the
 standard quantum limit. 
For example, the sensitivity of a generalized cat state converted from a thermal equilibrium state at finite temperature is found to be a few orders of magnitude better than the previous sensors, implying that the difficulty of state preparation could be drastically lifted.  Providing a wide class that includes such a peculiar state, our work paves the way to broaden the
 applications  of quantum metrology.

\begin{acknowledgments}
We thank R. Hamazaki and H. Hakoshima for discussions. 
M.T. was supported by the Japan Society for the Promotion of Science through Program for Leading Graduate Schools (ALPS) and a JSPS fellowship (JSPS KAKENHI Grant No. JP19J12884).
This work was supported by The Japan Society for the Promotion of Science, KAKENHI Grants No. 15H05700 and No. 19H01810,  MEXT KAKENHI Grant No. 15H05870, and CREST (Grant No. JPMJCR1774).
\end{acknowledgments}

\appendix
\begin{widetext}
\section{Derivation of (\ref{withdephasing}) }

If only a single qubit dephases, the Hamiltonian is
\begin{align}
\hat H_0 +\hat H_{\rm{int1}}(t),
\end{align}
where
\begin{align}
\hat H_0= \omega \sum_{l=1}^N \hat a(l) =\omega \hat A,\\
\hat H_{\rm{int1}}(t)=\lambda f_l(t)\hat a(l).
\end{align}
Since $[\hat H_0, \hat H_{\rm{int1}}(t)]=0$, the interaction picture is convenient:
\begin{align}
\hat \rho^I(t)= e^{i \hat H_0 t }\hat \rho(t) e^{-i \hat H_0 t},\\
\frac{d \hat \rho^I(t)}{dt}= -i [\hat H_{\rm{int1}}(t), \hat \rho^I(t)].
\end{align}
Then we have
\begin{align}
\hat \rho^I(t_{\rm int})
=\hat \rho(0) +\sum_{n=1}^\infty (-i\lambda)^n \int_0^{t_{\rm int}}\int_0^{t_1}\cdots \int_0^{t_{n-1}}dt_1 dt_2\cdots dt_n [\hat H_{\rm{int1}}(t_1),[\hat H_{\rm{int1}}(t_2),\cdots[\hat H_{\rm{int1}}(t_n),\hat \rho(0)]]].
\end{align}
Taking the average over the ensemble of the noise, we obtain
\begin{align}
\hat \rho^I(t_{\rm int})-\hat \rho(0)
&=\sum_{n=1}^\infty (-i\lambda)^n \overline{f_l(t_1)f_l(t_2)\cdots f_l(t_n)}\int_0^{t_{\rm int}}\int_0^{t_1}\cdots \int_0^{t_{n-1}}dt_1 dt_2\cdots dt_n [\hat a(l),\hat \rho(0)]_n.
\end{align}
Here, we define $[\hat O_1,\hat O_2]_k$ as $[\hat O_1,\hat O_2]_{k+1}=[\hat O_1,[\hat O_1,\hat O_2]_k]$ and $[\hat O_1,\hat O_2]_0= \hat O_2$.
Since we assume $\overline{f_j(t)f_k(t')}=\delta_{j,k}$ and  the $m(>2)$th cumulants are zero for Gaussian noise, $\overline{f_l(t_1)f_l(t_2)\cdots f_l(t_n)}$ can be decomposed into 
\begin{align}
\overline{f_l(t_1)f_l(t_2)\cdots f_l(t_{2n})}
&= \sum_{all\,combination}\overline{f(t'_1)f(t'_2)}\,\overline{f(t'_3)f(t'_4)}\cdots\overline{f(t'_{2n-1})f(t'_{2n})}\\
&=(2n-1)(2n-3)\cdots3\cdot 1=(2n-1)!!
\end{align}
and
\begin{align}
\overline{f_l(t_1)f_l(t_2)\cdots f_l(t_{2n+1})}
&= \sum_{all\,combination}\overline{f(t'_1)f(t'_2)}\,\overline{f(t'_3)f(t'_4)}\cdots\overline{f(t'_{2n-1})f(t'_{2n})}\,\overline{f(t'_{2n+1})}\\
&=0.
\end{align}
Therefore, we have
\begin{align}
\hat \rho^I(t_{\rm int})-\hat \rho(0)
&=\sum_{n=1}^\infty (-i\lambda)^{2n} (2n-1)!!\int_0^{t_{\rm int}}\int_0^{t_1}\cdots \int_0^{t_{n-1}}dt_1 dt_2\cdots dt_{2n} [\hat a(l),\hat \rho(0)]_{2n}\\
&=\sum_{n=1}^\infty (-\lambda^2)^{n} (2n-1)!!t_{\rm int}^n\frac{1}{(2n)!} [\hat a(l),\hat \rho(0)]_{2n}\\
&=\sum_{n=1}^\infty (-\lambda^2t_{\rm int})^{n} \frac{1}{2^nn!} [\hat a(l),\hat \rho(0)]_{2n}
\end{align}
By assuming $\hat a(l)^2=\hat 1$, which holds for $\pm \hat \sigma_{x,y,z}$, we can simplify the commuation:
\begin{align}
[\hat a(l),\hat \rho(0)]_{2n}= \frac{2^{2n}}{2}(\hat\rho(0)-\hat a(l)\hat \rho(0)\hat a(l)).
\end{align}
This gives us
\begin{align}
\hat \rho^I(t_{\rm int})-\hat \rho(0)
&=\sum_{n=1}^\infty (-\lambda^2t_{\rm int})^{n} \frac{1}{2^nn!} \frac{2^{2n}}{2}(\hat\rho(0)-\hat a(l)\hat \rho(0)\hat a(l))\\
&=\frac{1}{2}\sum_{n=1}^\infty \frac{(-2\lambda^2t_{\rm int})^{n}}{n!} (\hat\rho(0)-\hat a(l)\hat \rho(0)\hat a(l))\\
&=\frac{1}{2}\sum_{n=0}^\infty \frac{(-2\lambda^2t_{\rm int})^{n}}{n!} (\hat\rho(0)-\hat a(l)\hat \rho(0)\hat a(l))-(\frac{\hat\rho(0)-\hat a(l)\hat \rho(0)\hat a(l)}{2})\\
&=\frac{e^{-2\lambda^2t_{\rm int}}}{2}(\hat\rho(0)-\hat a(l)\hat \rho(0)\hat a(l))-\frac{\hat\rho(0)-\hat a(l)\hat \rho(0)\hat a(l)}{2},\\
\hat \rho^I(t_{\rm int})
&=\hat \rho(0)+\frac{e^{-2\lambda^2t_{\rm int}}-1}{2}\hat\rho(0)+\frac{1-e^{-2\lambda^2t_{\rm int}}}{2}\hat a(l)\hat \rho(0)\hat a(l)\\
&=\frac{1+e^{-2\lambda^2t_{\rm int}}}{2}\hat \rho(0)+\frac{1-e^{-2\lambda^2t_{\rm int}}}{2}\hat a(l)\hat \rho(0)\hat a(l).
\end{align}

When $N$ spins dephase, i.e., $\hat H_{\rm{int}}(t)=\sum_{l=1}^N\lambda f_l(t)\hat a(l)$, $\hat \rho^I(t_{\rm int})$ can be expressed as
\begin{align}
\hat\rho(t_{\rm int})^I= \epsilon_N(\epsilon_{N-1}\cdots\epsilon_1(\hat \rho(0))),
\end{align}
where
\begin{align}
\epsilon_j(\hat \rho(0))= \frac{1+e^{-\lambda^2t_{\rm int}^2}}{2}\hat \rho(0)+\frac{1-e^{\lambda^2t_{\rm int}^2}}{2}\hat a(j)\hat \rho(0) \hat a(j)
\end{align}
since $[\hat a(l),\hat a(k)]=0$ for arbitrary pair of $(l,k)$.
Explicitly expressing, we have
\begin{align}
&\rho(t_{\rm int})^I=
\left(\frac{1+e^{-\lambda^2t_{\rm int}^2}}{2}\right)^N\hat \rho(0)
+\left(\frac{1+e^{-\lambda^2t_{\rm int}^2}}{2}\right)^{N-1}\left(\frac{1-e^{-\lambda^2t_{\rm int}^2}}{2}\right)\sum_{j=1}^N\hat a(j) \hat \rho(0) \hat a(j)
+\cdots\nonumber\\
&+\left(\frac{1-e^{-\lambda^2t_{\rm int}^2}}{2}\right)^N\hat a_N\hat a_{N-1}\cdots \hat a_1\hat \rho(0) \hat a_1\cdots \hat a_{N-1}\hat a_N,\\
&\hat\rho(t)=
e^{-i\hat H_0t_{\rm int}}\left(\left(\frac{1+e^{-\lambda^2t_{\rm int}^2}}{2}\right)^N\hat \rho(0)
+\left(\frac{1+e^{-\lambda^2t_{\rm int}^2}}{2}\right)^{N-1}\left(\frac{1-e^{-\lambda^2t_{\rm int}^2}}{2}\right)\sum_{j=1}^N\hat a(j) \hat \rho(0) \hat a(j)
+\cdots\right.\nonumber\\
&\left.+\left(\frac{1-e^{-\lambda^2t_{\rm int}^2}}{2}\right)^N\hat a_N\hat a_{N-1}\cdots \hat a_1\hat \rho(0) \hat a_1\cdots \hat a_{N-1}\hat a_N\right)e^{i\hat H_0t_{\rm int}}
\end{align}
For 
\begin{align}
\hat \rho':
&=\hat\rho(t_{\rm int})-\left(\frac{1+e^{-\lambda^2t_{\rm int}^2}}{2}\right)^Ne^{-i\hat H_0t_{\rm int}}\hat\rho(0) e^{i\hat H_0t_{\rm int}}\\
&=e^{-i\hat H_0t_{\rm int}}\left(
\left(\frac{1+e^{-\lambda^2t_{\rm int}^2}}{2}\right)^{N-1}\left(\frac{1-e^{-\lambda^2t_{\rm int}^2}}{2}\right)\sum_{j=1}^N\hat a(j) \hat \rho(0) \hat a(j)
+\cdots\right.\nonumber\\
&\quad\left.+\left(\frac{1-e^{-\lambda^2t_{\rm int}^2}}{2}\right)^N\hat a_N\hat a_{N-1}\cdots \hat a_1\hat \rho(0) \hat a_1\cdots \hat a_{N-1}\hat a_N\right)e^{i\hat H_0t_{\rm int}}\\
&=:e^{-i\hat H_0t_{\rm int}}\hat \rho_0' e^{i\hat H_0t_{\rm int}},
\end{align}
we have
\begin{align}
\left|\frac{d\mathrm{Tr}(\hat \rho'\hat\eta)}{d\omega}\right|
&=\left|\frac{d}{d\omega}\sum_{k=0}^{\infty}\frac{(i\omega t_{\rm int})^k}{k!}\mathrm{Tr}(\rho_0'[\hat A,\hat\eta]_k)\right|\\
&\leq 
2\|\hat A\|t_{\rm int}e^{2\omega t_{\rm int}\|\hat A\|}\|\rho_0'\|\\
&=2\|\hat A\|t_{\rm int}e^{2\omega t_{\rm int}\|\hat A\|}\times\|\left(\frac{1+e^{-\lambda^2t_{\rm int}^2}}{2}\right)^{N-1}\left(\frac{1-e^{-\lambda^2t_{\rm int}^2}}{2}\right)\sum_{j=1}^N\hat a(j) \hat \rho(0) \hat a(j)+
\nonumber\\
&\quad\cdots
+\left(\frac{1-e^{-\lambda^2t_{\rm int}^2}}{2}\right)^N\hat a_N\hat a_{N-1}\cdots \hat a_1\hat \rho(0) \hat a_1\cdots \hat a_{N-1}\hat a_N\|\\
&=2\|\hat A\|t_{\rm int}e^{2\omega t_{\rm int}\|\hat A\|}\left(\left(\frac{1+e^{-\lambda^2t_{\rm int}^2}}{2}\right)^{N-1}\left(\frac{1-e^{-\lambda^2t_{\rm int}^2}}{2}\right)\binom{N}{1}
+\cdots
+\left(\frac{1-e^{-\lambda^2t_{\rm int}^2}}{2}\right)^N\binom{N}{N}\right)\\
&=2\|\hat A\|t_{\rm int}e^{2\omega t_{\rm int}\|\hat A\|}\left(1-\left(\frac{1+e^{-\lambda^2 t_{\rm int}^2}}{2}\right)^N\right).
\end{align}
Here, we used the following formulas:
\begin{align}
e^{i\omega \hat At_{\rm int}}\hat \eta e^{-i\omega \hat At_{\rm int}}= \sum_{k=0}^{\infty}\frac{(i\omega t_{\rm int})^k}{k!}[\hat A,\hat \eta]_k,\label{exptocom}\\
|\mathrm{Tr}(\hat\rho[\hat A,\hat \eta]_k)|\leq 2^k\|\hat A\|^k.
\end{align}

The derivation of (\ref{exptocom}) is as follows:
\begin{align}
e^{i\omega \hat At}\hat \eta e^{-i\omega \hat At}
&=
\sum_{m,m',\nu,\nu'}e^{i\omega \hat At}\ket{m,\nu}\bra{m,\nu}\hat \eta \ket{m',\nu'}\bra{m',\nu'}e^{-i\omega \hat At}\\
&=
\sum_{m,m',\nu,\nu'}e^{i\omega A_mt}\ket{m,\nu}\bra{m,\nu}\hat \eta \ket{m',\nu'}\bra{m',\nu'}e^{-i\omega A_{m'}t}\\
&=
\sum_{m,m',\nu,\nu'}e^{i\omega (A_m-A_{m'})t}\ket{m,\nu}\bra{m,\nu}\hat \eta \ket{m',\nu'}\bra{m',\nu'},\\
\end{align}
\begin{align}
\sum_{k=0}^{\infty}\frac{(i\omega t)^k}{k!}[\hat A,\hat \eta]_k
&=
\sum_{k=0}^{\infty}\frac{(i\omega t)^k}{k!}\sum_{m,m',\nu,\nu'}\ket{m,\nu}\bra{m,\nu}[\hat A,\hat \eta]_k\ket{m',\nu'}\bra{m',\nu'}\\
&=
\sum_{k=0}^{\infty}\frac{(i\omega t)^k}{k!}\sum_{m,m',\nu,\nu'}\ket{m,\nu}\sum_{k'=0}^{k}(-1)^{k'}\bra{m,\nu}\hat A^{k-k'}\hat \eta \hat A^{k'}\ket{m',\nu'}\bra{m',\nu'}\\
&=
\sum_{k=0}^{\infty}\frac{(i\omega t)^k}{k!}\sum_{m,m',\nu,\nu'}\ket{m,\nu}\sum_{k'=0}^{k}(-1)^{k'} A_m^{k-k'}A^{k'}_{m'}\bra{m,\nu}\hat \eta \ket{m',\nu'}\bra{m',\nu'}\\
&=
\sum_{k=0}^{\infty}\frac{(i\omega t)^k}{k!}\sum_{m,m',\nu,\nu'} (A_m-A_{m'})^k\ket{m,\nu}\bra{m,\nu}\hat \eta \ket{m',\nu'}\bra{m',\nu'}\\
&=
\sum_{m,m',\nu,\nu'} e^{i\omega t(A_m-A_{m'})}\ket{m,\nu}\bra{m,\nu}\hat \eta \ket{m',\nu'}\bra{m',\nu'},
\end{align}
where $\hat A\ket{m,\nu}=A_m\ket{m,\nu}$ and $\nu$ labels the degeneracy.

So the denominator of the sensitivity is
\begin{align}
\sqrt{T/t_{\rm int}}\left|\frac{d\mathrm{Tr}(\hat \eta \hat\rho(t_{\rm int}))}{d\omega}\right|
&\geq
\sqrt{T/t_{\rm int}}(\left|\frac{d\mathrm{Tr}(\hat\eta e^{-i\omega \hat At_{\rm int}}\hat\rho(0)e^{i\omega \hat At_{\rm int}})}{d\omega}\right|\left(\frac{1+e^{-\lambda^2t_{\rm int}^2}}{2}\right)^N
-\left|\frac{d\mathrm{Tr}(\hat\eta \hat\rho')}{d\omega}\right|)\\
&\geq
\sqrt{T/t_{\rm int}}(\left|\frac{dP}{d\omega}\right| \left(\frac{1+e^{-\lambda^2t_{\rm int}^2}}{2}\right)^N 
-2\|\hat A\|t_{\rm int}e^{2\omega t_{\rm int}\|\hat A\|}\left(1-\left(\frac{1+e^{-\lambda^2 t_{\rm int}^2}}{2}\right)^N\right),
\end{align}
where
\begin{align}
\left|\frac{dP}{d\omega}\right|
\geq \left| \left|\omega t_{\rm int}^2\mathrm{Tr}(\hat \rho(0)[\hat A,[\hat A,\hat\eta]])\right|
-\left|it_{\rm int}\mathrm{Tr}(\hat \rho(0) [\hat A,\hat \eta])\right|
\right|
-2t_{\rm int}\|\hat A\|(e^{2\omega t_{\rm int}\|\hat A\|}-1-2\omega t_{\rm int}\|\hat A\|).
\end{align}
Using the result of the  case where there is no noise, we obtain (\ref{withdephasing})
\begin{align}
\delta \omega_{\rm deph}\sqrt T 
& \leq
(N\sqrt{t_{\rm int}})^{-1}
\Big[ p_1 p_2^2 
\Big(\frac{1+e^{-2\lambda^2t_{\rm int}^2}}{2}\Big)^N
\nonumber\\
&-2e^{2\omega t_{\rm int} \|\hat A\|
}\frac{\|\hat A\|}{N} \Big(1-\Big(\frac{1+e^{-2\lambda^2t_{\rm int}^2}}{2}\Big)^N \Big) \Big]^{-1}.
\end{align}

\section{The scaling of the uncertainty of the estimation 
}
In the standard setup of the quantum metrology, generalized cat states always give the scalings either $\delta \omega =\Theta(N^{-3/4})$ with a finite dephasing rate or  $\delta \omega =\Theta(N^{-1} )$ with a zero dephasing rate.
(For convenience,  we express the uncertainty as $\delta\omega$ regardless of the existence of dephasing in this section.)
 We do not obtain the intermediate scaling such as $\delta \omega =\Theta(N^k)$ with $-1<k<-3/4$ even with a small dephasing.
In this section, we explain the reason  by considering a GHZ state $\frac{1}{\sqrt{2}} (|0\rangle^{\otimes N} +|1\rangle^{\otimes N} ) $ of $N$ qubits as an example.

When we try to estimate $\omega$ of $\hat H=\sum_{j=1}^{N}\frac{\omega }{2}\hat{\sigma }_z^{(j)}$,
we (1) prepare the GHZ state, (2) let the state evolve for time $t_{\rm int}$, (3) read out, and (4) repeat from (1) to (3) for $T/t_{\rm int}$ times (assuming the state preparation and the readout are done instantaneously).
Here, $T$ is the total measurement time which we can freely fix at some finite value.
In the presence of non-Markovian dephasing, 
the uncertainty $\delta\omega$ is calculated as \cite{matsuzaki2011magnetic}
\begin{eqnarray}
 \delta \omega  = 
\frac{e^{\frac{Nt_{\rm int}^2}{(T_2)^2}}}{N\sqrt{Tt_{\rm int}}},\label{ghzdw}
\end{eqnarray}
where $T_2$ is the coherence time  of a single qubit  determined by the physical system.
Our aim is to minimize $\delta \omega\sqrt{T}$  by tuning $t_{\rm int}$, and to see how it scales with $N$.

For finite $T_2$,  $\delta\omega \sqrt{T}$ has the minimum value $\frac{\sqrt{2}\exp(1/4)}{N^{3/4}\sqrt{T_2}}$ at $t_{\rm int}=T_2/2\sqrt{N}$.
As we can see from Fig.~\ref{twoscalings}, the minimum value of $\delta\omega \sqrt{T}$ moves to the right as $T_2$ increases.
In the limit of no dephasing, i.e., $T_2\rightarrow \infty$, $\delta\omega\sqrt{T}$ no longer has a minimum value. Instead, we find $\delta\omega\sqrt{T}\rightarrow \frac{1}{N\sqrt{t_{\rm int}}}$, which scales as $N^{-1}$ for $t_{\rm int}=\Theta(N^0)$.

\begin{figure}
 \centering
 \includegraphics[keepaspectratio, scale=0.5]
      {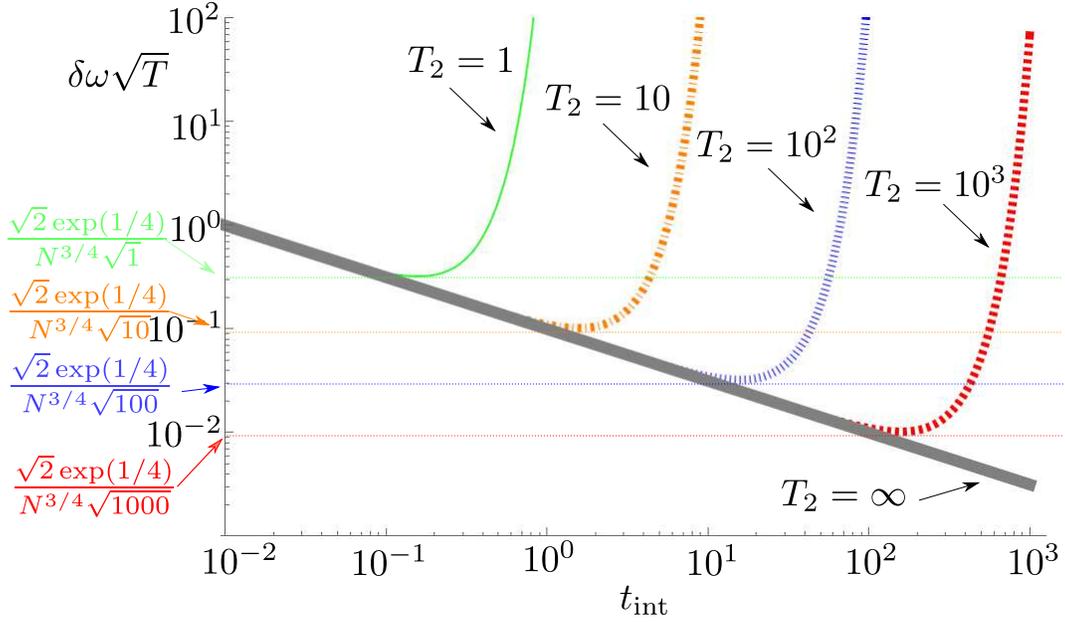}
 \caption{
Log-log plot of $\delta\omega \sqrt{T}$ against $t_{\rm int}$ for $N=10$. From the left, green, orange (dot-dashed), blue (dotted), and red (dashed) curves correspond to $T_2=1$, $T_2=10$, $T_2=10^2$, $T_2=10^3$, respectively. The gray (thick) line corresponds to  $T_2\rightarrow \infty$.
The minimum value varies in accordance with $T_2$, but it always scales as $N^{-3/4}$ as long as $T_2$ is finite. 
However, when $T_2\rightarrow \infty$, $\delta\omega\sqrt{T}$ takes the form $1/N\sqrt{t_{\rm int}}$ and keeps decreasing (without minimum values), giving another scaling $N^{-1}$ for the optimal uncertainty. } 
\label{twoscalings}
\end{figure}

The intuitive reason why $\delta\omega\sqrt{T}$ has a minimum value with $T_2<\infty$ is that 
while larger $t_{\rm int}$ gives more phase accumulation, which contributes to a better sensitivity, 
the amplitude of the state maintaining useful coherence for sensing
 diminishes with the increase of $t_{\rm int}$ because of the noise.
When there is no noise, on the other hand,  the latter does not occur. Hence the sensitivity keeps improving with the increase of $t_{\rm int}$  when  $T_2\rightarrow \infty$.




Although we describe the case with the GHZ state as an example, the same conclusion can be drawn with the field sensor with the generalized cat states.

Therefore, for the reason described above, we do not obtain the intermediate scaling such as $\delta \omega =\Theta(N^k)$ with $-1<k<-3/4$.

\section{Construction of $\hat \eta$}
In this section, we explain how to judge whether a given state is useful in metrology and show how to construct a projection $\hat \eta$ for a given state.
For an arbitrary $\hat \rho$, we can judge whether it is 
helpful in sensing $\omega$ of $\omega\hat A$
as follows:
Find the eigenvalue and eigenstate of $[\hat A, [\hat A, \hat \rho]]$.
If the sum of the positive eigenvalues is $\Theta(N^2)$, then it is a generalized cat state of $\hat A$, i.e., there exists a projection operator satisfying $\mathrm{Tr}(\hat \rho [\hat A,[\hat A,\hat \eta]])=\Theta(N^2)$.

The projection operator $\hat \eta$ for the Ramsey-type measurement with the 
ultimate
scaling can be constructed using the  eigenstates:
\begin{align}
\hat \eta= \sum_{e_n>0}\ket{n}\bra{n},
\end{align}
 where $\hat \rho \ket{n}=e_n\ket{n}$.

Let us give an example.
Let $\ket{\psi_{\lambda}}$ be the following state similar to the Schr\"odinger's cat state, but differs by the $\lambda$th spin,
\begin{align}
\ket{\psi_{\lambda}}:=
\frac{1}{\sqrt{2}}\ket{\downarrow}^{\otimes (\lambda-1)}\ket{\uparrow}\ket{\downarrow}^{\otimes (N-\lambda)}
+\frac{1}{\sqrt{2}}\ket{\uparrow}^{\otimes (\lambda-1)}\ket{\downarrow}\ket{\uparrow}^{\otimes (N-\lambda)}\quad(\lambda=1,2,...,N)
\end{align}
Then, let $\hat \rho_{ex}$ be a mixed state of $\ket{\psi_{\lambda}}$'s:
\begin{align}
\hat \rho_{ex}:=\frac{1}{N}\sum_{\lambda=1}^N\ket{\psi_{\lambda}}\bra{\psi_{\lambda}}.
\end{align}
The eigenstates with positive eigenvalues of $[\hat M_z,[\hat M_z, \hat \rho_{ex}]]$ are $\ket{\psi_{\lambda}}$'s,
and the sum of the eigenvalues is $2(N-2)^2$.
Hence the mixed state $\hat \rho_{ex}$ can be proven to achieve the ultimate scaling  in measuring $\hat M_z $ with a projection $\hat \eta=N\hat \rho$ after Ramsey-type protocol.

\section{Derivation of upper bound}
A numerical upper bound of $\delta\omega_{\rm deph}\sqrt{T}$
is obtained through calculating $\mathrm{Tr}(\hat\rho [\hat A,\hat \eta])$ and $\mathrm{Tr}(\hat\rho[\hat A,[\hat A,\hat\eta]])$ numerically, and then minimizing 
\begin{align}
(N\sqrt{t_{\rm int}})^{-1}
\Big[ U 
\Big(\frac{1+e^{-2\lambda^2t_{\rm int}^2}}{2}\Big)^N
-2e^{2\omega t_{\rm int} \|\hat A\|
} \Big(1-\Big(\frac{1+e^{-2\lambda^2t_{\rm int}^2}}{2}\Big)^N \Big) \Big]^{-1}
\label{upperbound}
\end{align}
by tuning $t_{\rm int}$,
where
\begin{align}
U:=
\left|\frac{ \left|\omega t_{\rm int}\mathrm{Tr}(\hat\rho[\hat A,[\hat A,\hat\eta]])\right|}{N}
-\frac{\left|i\mathrm{Tr}(\hat\rho [\hat A,\hat \eta])\right|}{N}
\right|
-2\frac{\|\hat A\|}{N}(e^{2\omega t_{\rm int}\|\hat A\|}-1-2\omega t_{\rm int}\|\hat A\|).
\end{align}
We then find $t_{\rm int}\propto 1/\sqrt N$ gives the optimal uncertainty $\delta\omega_{\rm deph}=\Theta(N^{3/4})$.

\section{Relation between QFI and $q$}
We would also like to comment that we revealed the unknown general relation between QFI and $q$.
Fr\"owis and D\"ur claim that the QFI
can characterize the macroscopicity of quantum states \cite{frowis2012measures,frowis2018macroscopic}; if the QFI
is of the order of $\Theta (N^2)$, they consider the quantum state as  macroscopic.
The relationship between QFI and $q$ for general
mixed states was an open question. 
Here, we showed $1/\sqrt {\rm {QFI}}\leq \delta\omega \leq \Theta(N^{-1})$ for $q=2$ states, assuring the lower bound of QFI to be large.
Connecting two criteria defined from different aspects, our results contribute to the further understanding
of
physics.
\end{widetext}

\bibliography{ref}

\begin{thebibliography}{63}%
\makeatletter
\providecommand \@ifxundefined [1]{%
 \@ifx{#1\undefined}
}%
\providecommand \@ifnum [1]{%
 \ifnum #1\expandafter \@firstoftwo
 \else \expandafter \@secondoftwo
 \fi
}%
\providecommand \@ifx [1]{%
 \ifx #1\expandafter \@firstoftwo
 \else \expandafter \@secondoftwo
 \fi
}%
\providecommand \natexlab [1]{#1}%
\providecommand \enquote  [1]{``#1''}%
\providecommand \bibnamefont  [1]{#1}%
\providecommand \bibfnamefont [1]{#1}%
\providecommand \citenamefont [1]{#1}%
\providecommand \href@noop [0]{\@secondoftwo}%
\providecommand \href [0]{\begingroup \@sanitize@url \@href}%
\providecommand \@href[1]{\@@startlink{#1}\@@href}%
\providecommand \@@href[1]{\endgroup#1\@@endlink}%
\providecommand \@sanitize@url [0]{\catcode `\\12\catcode `\$12\catcode
  `\&12\catcode `\#12\catcode `\^12\catcode `\_12\catcode `\%12\relax}%
\providecommand \@@startlink[1]{}%
\providecommand \@@endlink[0]{}%
\providecommand \url  [0]{\begingroup\@sanitize@url \@url }%
\providecommand \@url [1]{\endgroup\@href {#1}{\urlprefix }}%
\providecommand \urlprefix  [0]{URL }%
\providecommand \Eprint [0]{\href }%
\providecommand \doibase [0]{http://dx.doi.org/}%
\providecommand \selectlanguage [0]{\@gobble}%
\providecommand \bibinfo  [0]{\@secondoftwo}%
\providecommand \bibfield  [0]{\@secondoftwo}%
\providecommand \translation [1]{[#1]}%
\providecommand \BibitemOpen [0]{}%
\providecommand \bibitemStop [0]{}%
\providecommand \bibitemNoStop [0]{.\EOS\space}%
\providecommand \EOS [0]{\spacefactor3000\relax}%
\providecommand \BibitemShut  [1]{\csname bibitem#1\endcsname}%
\let\auto@bib@innerbib\@empty
\bibitem [{\citenamefont {Giovannetti}\ \emph {et~al.}(2004)\citenamefont
  {Giovannetti}, \citenamefont {Lloyd},\ and\ \citenamefont
  {Maccone}}]{giovannetti2004quantum}%
  \BibitemOpen
  \bibfield  {author} {\bibinfo {author} {\bibfnamefont {V.}~\bibnamefont
  {Giovannetti}}, \bibinfo {author} {\bibfnamefont {S.}~\bibnamefont {Lloyd}},
  \ and\ \bibinfo {author} {\bibfnamefont {L.}~\bibnamefont {Maccone}},\
  }\href@noop {} {\bibfield  {journal} {\bibinfo  {journal} {Science}\ }\textbf
  {\bibinfo {volume} {306}},\ \bibinfo {pages} {1330} (\bibinfo {year}
  {2004})}\BibitemShut {NoStop}%
\bibitem [{\citenamefont {Giovannetti}\ \emph {et~al.}(2011)\citenamefont
  {Giovannetti}, \citenamefont {Lloyd},\ and\ \citenamefont
  {Maccone}}]{giovannetti2011advances}%
  \BibitemOpen
  \bibfield  {author} {\bibinfo {author} {\bibfnamefont {V.}~\bibnamefont
  {Giovannetti}}, \bibinfo {author} {\bibfnamefont {S.}~\bibnamefont {Lloyd}},
  \ and\ \bibinfo {author} {\bibfnamefont {L.}~\bibnamefont {Maccone}},\
  }\href@noop {} {\bibfield  {journal} {\bibinfo  {journal} {Nature photonics}\
  }\textbf {\bibinfo {volume} {5}},\ \bibinfo {pages} {222} (\bibinfo {year}
  {2011})}\BibitemShut {NoStop}%
\bibitem [{\citenamefont {Taylor}\ and\ \citenamefont
  {Bowen}(2016)}]{taylor2016quantum}%
  \BibitemOpen
  \bibfield  {author} {\bibinfo {author} {\bibfnamefont {M.~A.}\ \bibnamefont
  {Taylor}}\ and\ \bibinfo {author} {\bibfnamefont {W.~P.}\ \bibnamefont
  {Bowen}},\ }\href@noop {} {\bibfield  {journal} {\bibinfo  {journal} {Physics
  Reports}\ }\textbf {\bibinfo {volume} {615}},\ \bibinfo {pages} {1} (\bibinfo
  {year} {2016})}\BibitemShut {NoStop}%
\bibitem [{\citenamefont {Degen}\ \emph {et~al.}(2017)\citenamefont {Degen},
  \citenamefont {Reinhard},\ and\ \citenamefont
  {Cappellaro}}]{degen2017quantum}%
  \BibitemOpen
  \bibfield  {author} {\bibinfo {author} {\bibfnamefont {C.~L.}\ \bibnamefont
  {Degen}}, \bibinfo {author} {\bibfnamefont {F.}~\bibnamefont {Reinhard}}, \
  and\ \bibinfo {author} {\bibfnamefont {P.}~\bibnamefont {Cappellaro}},\
  }\href@noop {} {\bibfield  {journal} {\bibinfo  {journal} {Reviews of modern
  physics}\ }\textbf {\bibinfo {volume} {89}},\ \bibinfo {pages} {035002}
  (\bibinfo {year} {2017})}\BibitemShut {NoStop}%
\bibitem [{\citenamefont {Wineland}\ \emph {et~al.}(1992)\citenamefont
  {Wineland}, \citenamefont {Bollinger}, \citenamefont {Itano}, \citenamefont
  {Moore},\ and\ \citenamefont {Heinzen}}]{wineland1992spin}%
  \BibitemOpen
  \bibfield  {author} {\bibinfo {author} {\bibfnamefont {D.~J.}\ \bibnamefont
  {Wineland}}, \bibinfo {author} {\bibfnamefont {J.~J.}\ \bibnamefont
  {Bollinger}}, \bibinfo {author} {\bibfnamefont {W.~M.}\ \bibnamefont
  {Itano}}, \bibinfo {author} {\bibfnamefont {F.~L.}\ \bibnamefont {Moore}}, \
  and\ \bibinfo {author} {\bibfnamefont {D.~J.}\ \bibnamefont {Heinzen}},\
  }\href@noop {} {\bibfield  {journal} {\bibinfo  {journal} {Physical Review
  A}\ }\textbf {\bibinfo {volume} {46}},\ \bibinfo {pages} {R6797} (\bibinfo
  {year} {1992})}\BibitemShut {NoStop}%
\bibitem [{\citenamefont {Wineland}\ \emph {et~al.}(1994)\citenamefont
  {Wineland}, \citenamefont {Bollinger}, \citenamefont {Itano},\ and\
  \citenamefont {Heinzen}}]{wineland1994squeezed}%
  \BibitemOpen
  \bibfield  {author} {\bibinfo {author} {\bibfnamefont {D.~J.}\ \bibnamefont
  {Wineland}}, \bibinfo {author} {\bibfnamefont {J.~J.}\ \bibnamefont
  {Bollinger}}, \bibinfo {author} {\bibfnamefont {W.~M.}\ \bibnamefont
  {Itano}}, \ and\ \bibinfo {author} {\bibfnamefont {D.~J.}\ \bibnamefont
  {Heinzen}},\ }\href@noop {} {\bibfield  {journal} {\bibinfo  {journal}
  {Physical Review A}\ }\textbf {\bibinfo {volume} {50}},\ \bibinfo {pages}
  {67} (\bibinfo {year} {1994})}\BibitemShut {NoStop}%
\bibitem [{\citenamefont {T{\'o}th}\ and\ \citenamefont
  {Apellaniz}(2014)}]{toth2014quantum}%
  \BibitemOpen
  \bibfield  {author} {\bibinfo {author} {\bibfnamefont {G.}~\bibnamefont
  {T{\'o}th}}\ and\ \bibinfo {author} {\bibfnamefont {I.}~\bibnamefont
  {Apellaniz}},\ }\href@noop {} {\bibfield  {journal} {\bibinfo  {journal}
  {Journal of Physics A: Mathematical and Theoretical}\ }\textbf {\bibinfo
  {volume} {47}},\ \bibinfo {pages} {424006} (\bibinfo {year}
  {2014})}\BibitemShut {NoStop}%
\bibitem [{\citenamefont {Le~Sage}\ \emph {et~al.}(2013)\citenamefont
  {Le~Sage}, \citenamefont {Arai}, \citenamefont {Glenn}, \citenamefont
  {DeVience}, \citenamefont {Pham}, \citenamefont {Rahn-Lee}, \citenamefont
  {Lukin}, \citenamefont {Yacoby}, \citenamefont {Komeili},\ and\ \citenamefont
  {Walsworth}}]{le2013optical}%
  \BibitemOpen
  \bibfield  {author} {\bibinfo {author} {\bibfnamefont {D.}~\bibnamefont
  {Le~Sage}}, \bibinfo {author} {\bibfnamefont {K.}~\bibnamefont {Arai}},
  \bibinfo {author} {\bibfnamefont {D.}~\bibnamefont {Glenn}}, \bibinfo
  {author} {\bibfnamefont {S.}~\bibnamefont {DeVience}}, \bibinfo {author}
  {\bibfnamefont {L.}~\bibnamefont {Pham}}, \bibinfo {author} {\bibfnamefont
  {L.}~\bibnamefont {Rahn-Lee}}, \bibinfo {author} {\bibfnamefont
  {M.}~\bibnamefont {Lukin}}, \bibinfo {author} {\bibfnamefont
  {A.}~\bibnamefont {Yacoby}}, \bibinfo {author} {\bibfnamefont
  {A.}~\bibnamefont {Komeili}}, \ and\ \bibinfo {author} {\bibfnamefont
  {R.}~\bibnamefont {Walsworth}},\ }\href@noop {} {\bibfield  {journal}
  {\bibinfo  {journal} {Nature}\ }\textbf {\bibinfo {volume} {496}},\ \bibinfo
  {pages} {486} (\bibinfo {year} {2013})}\BibitemShut {NoStop}%
\bibitem [{\citenamefont {Paris}(2009)}]{paris2009quantum}%
  \BibitemOpen
  \bibfield  {author} {\bibinfo {author} {\bibfnamefont {M.~G.}\ \bibnamefont
  {Paris}},\ }\href@noop {} {\bibfield  {journal} {\bibinfo  {journal}
  {International Journal of Quantum Information}\ }\textbf {\bibinfo {volume}
  {7}},\ \bibinfo {pages} {125} (\bibinfo {year} {2009})}\BibitemShut {NoStop}%
\bibitem [{\citenamefont {Chin}\ \emph {et~al.}(2012)\citenamefont {Chin},
  \citenamefont {Huelga},\ and\ \citenamefont {Plenio}}]{chin2012quantum}%
  \BibitemOpen
  \bibfield  {author} {\bibinfo {author} {\bibfnamefont {A.~W.}\ \bibnamefont
  {Chin}}, \bibinfo {author} {\bibfnamefont {S.~F.}\ \bibnamefont {Huelga}}, \
  and\ \bibinfo {author} {\bibfnamefont {M.~B.}\ \bibnamefont {Plenio}},\
  }\href@noop {} {\bibfield  {journal} {\bibinfo  {journal} {Physical review
  letters}\ }\textbf {\bibinfo {volume} {109}},\ \bibinfo {pages} {233601}
  (\bibinfo {year} {2012})}\BibitemShut {NoStop}%
\bibitem [{\citenamefont {Chaves}\ \emph {et~al.}(2013)\citenamefont {Chaves},
  \citenamefont {Brask}, \citenamefont {Markiewicz}, \citenamefont
  {Ko{\l}ody{\'n}ski},\ and\ \citenamefont {Ac{\'\i}n}}]{chaves2013noisy}%
  \BibitemOpen
  \bibfield  {author} {\bibinfo {author} {\bibfnamefont {R.}~\bibnamefont
  {Chaves}}, \bibinfo {author} {\bibfnamefont {J.}~\bibnamefont {Brask}},
  \bibinfo {author} {\bibfnamefont {M.}~\bibnamefont {Markiewicz}}, \bibinfo
  {author} {\bibfnamefont {J.}~\bibnamefont {Ko{\l}ody{\'n}ski}}, \ and\
  \bibinfo {author} {\bibfnamefont {A.}~\bibnamefont {Ac{\'\i}n}},\ }\href@noop
  {} {\bibfield  {journal} {\bibinfo  {journal} {Physical review letters}\
  }\textbf {\bibinfo {volume} {111}},\ \bibinfo {pages} {120401} (\bibinfo
  {year} {2013})}\BibitemShut {NoStop}%
\bibitem [{\citenamefont {Jones}\ \emph {et~al.}(2009)\citenamefont {Jones},
  \citenamefont {Karlen}, \citenamefont {Fitzsimons}, \citenamefont {Ardavan},
  \citenamefont {Benjamin}, \citenamefont {Briggs},\ and\ \citenamefont
  {Morton}}]{Jones1166}%
  \BibitemOpen
  \bibfield  {author} {\bibinfo {author} {\bibfnamefont {J.~A.}\ \bibnamefont
  {Jones}}, \bibinfo {author} {\bibfnamefont {S.~D.}\ \bibnamefont {Karlen}},
  \bibinfo {author} {\bibfnamefont {J.}~\bibnamefont {Fitzsimons}}, \bibinfo
  {author} {\bibfnamefont {A.}~\bibnamefont {Ardavan}}, \bibinfo {author}
  {\bibfnamefont {S.~C.}\ \bibnamefont {Benjamin}}, \bibinfo {author}
  {\bibfnamefont {G.~A.~D.}\ \bibnamefont {Briggs}}, \ and\ \bibinfo {author}
  {\bibfnamefont {J.~J.~L.}\ \bibnamefont {Morton}},\ }\href {\doibase
  10.1126/science.1170730} {\bibfield  {journal} {\bibinfo  {journal}
  {Science}\ }\textbf {\bibinfo {volume} {324}},\ \bibinfo {pages} {1166}
  (\bibinfo {year} {2009})}\BibitemShut {NoStop}%
\bibitem [{\citenamefont {Huelga}\ \emph {et~al.}(1997)\citenamefont {Huelga},
  \citenamefont {Macchiavello}, \citenamefont {Pellizzari}, \citenamefont
  {Ekert}, \citenamefont {Plenio},\ and\ \citenamefont
  {Cirac}}]{huelga1997improvement}%
  \BibitemOpen
  \bibfield  {author} {\bibinfo {author} {\bibfnamefont {S.~F.}\ \bibnamefont
  {Huelga}}, \bibinfo {author} {\bibfnamefont {C.}~\bibnamefont
  {Macchiavello}}, \bibinfo {author} {\bibfnamefont {T.}~\bibnamefont
  {Pellizzari}}, \bibinfo {author} {\bibfnamefont {A.~K.}\ \bibnamefont
  {Ekert}}, \bibinfo {author} {\bibfnamefont {M.~B.}\ \bibnamefont {Plenio}}, \
  and\ \bibinfo {author} {\bibfnamefont {J.~I.}\ \bibnamefont {Cirac}},\
  }\href@noop {} {\bibfield  {journal} {\bibinfo  {journal} {Physical Review
  Letters}\ }\textbf {\bibinfo {volume} {79}},\ \bibinfo {pages} {3865}
  (\bibinfo {year} {1997})}\BibitemShut {NoStop}%
\bibitem [{\citenamefont {Kuzmich}\ \emph {et~al.}(1998)\citenamefont
  {Kuzmich}, \citenamefont {Bigelow},\ and\ \citenamefont
  {Mandel}}]{kuzmich1998atomic}%
  \BibitemOpen
  \bibfield  {author} {\bibinfo {author} {\bibfnamefont {A.}~\bibnamefont
  {Kuzmich}}, \bibinfo {author} {\bibfnamefont {N.}~\bibnamefont {Bigelow}}, \
  and\ \bibinfo {author} {\bibfnamefont {L.}~\bibnamefont {Mandel}},\
  }\href@noop {} {\bibfield  {journal} {\bibinfo  {journal} {EPL (Europhysics
  Letters)}\ }\textbf {\bibinfo {volume} {42}},\ \bibinfo {pages} {481}
  (\bibinfo {year} {1998})}\BibitemShut {NoStop}%
\bibitem [{\citenamefont {Fleischhauer}\ \emph {et~al.}(2000)\citenamefont
  {Fleischhauer}, \citenamefont {Matsko},\ and\ \citenamefont
  {Scully}}]{fleischhauer2000quantum}%
  \BibitemOpen
  \bibfield  {author} {\bibinfo {author} {\bibfnamefont {M.}~\bibnamefont
  {Fleischhauer}}, \bibinfo {author} {\bibfnamefont {A.~B.}\ \bibnamefont
  {Matsko}}, \ and\ \bibinfo {author} {\bibfnamefont {M.~O.}\ \bibnamefont
  {Scully}},\ }\href@noop {} {\bibfield  {journal} {\bibinfo  {journal}
  {Physical Review A}\ }\textbf {\bibinfo {volume} {62}},\ \bibinfo {pages}
  {013808} (\bibinfo {year} {2000})}\BibitemShut {NoStop}%
\bibitem [{\citenamefont {Geremia}\ \emph {et~al.}(2003)\citenamefont
  {Geremia}, \citenamefont {Stockton}, \citenamefont {Doherty},\ and\
  \citenamefont {Mabuchi}}]{geremia2003quantum}%
  \BibitemOpen
  \bibfield  {author} {\bibinfo {author} {\bibfnamefont {J.~M.}\ \bibnamefont
  {Geremia}}, \bibinfo {author} {\bibfnamefont {J.~K.}\ \bibnamefont
  {Stockton}}, \bibinfo {author} {\bibfnamefont {A.~C.}\ \bibnamefont
  {Doherty}}, \ and\ \bibinfo {author} {\bibfnamefont {H.}~\bibnamefont
  {Mabuchi}},\ }\href@noop {} {\bibfield  {journal} {\bibinfo  {journal}
  {Physical review letters}\ }\textbf {\bibinfo {volume} {91}},\ \bibinfo
  {pages} {250801} (\bibinfo {year} {2003})}\BibitemShut {NoStop}%
\bibitem [{\citenamefont {Leibfried}\ \emph {et~al.}(2004)\citenamefont
  {Leibfried}, \citenamefont {Barrett}, \citenamefont {Schaetz}, \citenamefont
  {Britton}, \citenamefont {Chiaverini}, \citenamefont {Itano}, \citenamefont
  {Jost}, \citenamefont {Langer},\ and\ \citenamefont
  {Wineland}}]{leibfried2004toward}%
  \BibitemOpen
  \bibfield  {author} {\bibinfo {author} {\bibfnamefont {D.}~\bibnamefont
  {Leibfried}}, \bibinfo {author} {\bibfnamefont {M.~D.}\ \bibnamefont
  {Barrett}}, \bibinfo {author} {\bibfnamefont {T.}~\bibnamefont {Schaetz}},
  \bibinfo {author} {\bibfnamefont {J.}~\bibnamefont {Britton}}, \bibinfo
  {author} {\bibfnamefont {J.}~\bibnamefont {Chiaverini}}, \bibinfo {author}
  {\bibfnamefont {W.~M.}\ \bibnamefont {Itano}}, \bibinfo {author}
  {\bibfnamefont {J.~D.}\ \bibnamefont {Jost}}, \bibinfo {author}
  {\bibfnamefont {C.}~\bibnamefont {Langer}}, \ and\ \bibinfo {author}
  {\bibfnamefont {D.~J.}\ \bibnamefont {Wineland}},\ }\href@noop {} {\bibfield
  {journal} {\bibinfo  {journal} {Science}\ }\textbf {\bibinfo {volume}
  {304}},\ \bibinfo {pages} {1476} (\bibinfo {year} {2004})}\BibitemShut
  {NoStop}%
\bibitem [{\citenamefont {Auzinsh}\ \emph {et~al.}(2004)\citenamefont
  {Auzinsh}, \citenamefont {Budker}, \citenamefont {Kimball}, \citenamefont
  {Rochester}, \citenamefont {Stalnaker}, \citenamefont {Sushkov},\ and\
  \citenamefont {Yashchuk}}]{auzinsh2004can}%
  \BibitemOpen
  \bibfield  {author} {\bibinfo {author} {\bibfnamefont {M.}~\bibnamefont
  {Auzinsh}}, \bibinfo {author} {\bibfnamefont {D.}~\bibnamefont {Budker}},
  \bibinfo {author} {\bibfnamefont {D.~F.}\ \bibnamefont {Kimball}}, \bibinfo
  {author} {\bibfnamefont {S.~M.}\ \bibnamefont {Rochester}}, \bibinfo {author}
  {\bibfnamefont {J.~E.}\ \bibnamefont {Stalnaker}}, \bibinfo {author}
  {\bibfnamefont {A.~O.}\ \bibnamefont {Sushkov}}, \ and\ \bibinfo {author}
  {\bibfnamefont {V.~V.}\ \bibnamefont {Yashchuk}},\ }\href@noop {} {\bibfield
  {journal} {\bibinfo  {journal} {Physical review letters}\ }\textbf {\bibinfo
  {volume} {93}},\ \bibinfo {pages} {173002} (\bibinfo {year}
  {2004})}\BibitemShut {NoStop}%
\bibitem [{\citenamefont {Dunningham}(2006)}]{dunningham2006using}%
  \BibitemOpen
  \bibfield  {author} {\bibinfo {author} {\bibfnamefont {J.~A.}\ \bibnamefont
  {Dunningham}},\ }\href@noop {} {\bibfield  {journal} {\bibinfo  {journal}
  {Contemporary Physics}\ }\textbf {\bibinfo {volume} {47}},\ \bibinfo {pages}
  {257} (\bibinfo {year} {2006})}\BibitemShut {NoStop}%
\bibitem [{\citenamefont {Matsuzaki}\ \emph {et~al.}(2011)\citenamefont
  {Matsuzaki}, \citenamefont {Benjamin},\ and\ \citenamefont
  {Fitzsimons}}]{matsuzaki2011magnetic}%
  \BibitemOpen
  \bibfield  {author} {\bibinfo {author} {\bibfnamefont {Y.}~\bibnamefont
  {Matsuzaki}}, \bibinfo {author} {\bibfnamefont {S.~C.}\ \bibnamefont
  {Benjamin}}, \ and\ \bibinfo {author} {\bibfnamefont {J.}~\bibnamefont
  {Fitzsimons}},\ }\href@noop {} {\bibfield  {journal} {\bibinfo  {journal}
  {Physical Review A}\ }\textbf {\bibinfo {volume} {84}},\ \bibinfo {pages}
  {012103} (\bibinfo {year} {2011})}\BibitemShut {NoStop}%
\bibitem [{\citenamefont {Demkowicz-Dobrza{\'n}ski}\ \emph
  {et~al.}(2012)\citenamefont {Demkowicz-Dobrza{\'n}ski}, \citenamefont
  {Ko{\l}ody{\'n}ski},\ and\ \citenamefont
  {Gu{\c{t}}{\u{a}}}}]{demkowicz2012elusive}%
  \BibitemOpen
  \bibfield  {author} {\bibinfo {author} {\bibfnamefont {R.}~\bibnamefont
  {Demkowicz-Dobrza{\'n}ski}}, \bibinfo {author} {\bibfnamefont
  {J.}~\bibnamefont {Ko{\l}ody{\'n}ski}}, \ and\ \bibinfo {author}
  {\bibfnamefont {M.}~\bibnamefont {Gu{\c{t}}{\u{a}}}},\ }\href@noop {}
  {\bibfield  {journal} {\bibinfo  {journal} {Nature communications}\ }\textbf
  {\bibinfo {volume} {3}},\ \bibinfo {pages} {1063} (\bibinfo {year}
  {2012})}\BibitemShut {NoStop}%
\bibitem [{\citenamefont {Bohnet}\ \emph {et~al.}(2014)\citenamefont {Bohnet},
  \citenamefont {Cox}, \citenamefont {Norcia}, \citenamefont {Weiner},
  \citenamefont {Chen},\ and\ \citenamefont {Thompson}}]{bohnet2014reduced}%
  \BibitemOpen
  \bibfield  {author} {\bibinfo {author} {\bibfnamefont {J.~G.}\ \bibnamefont
  {Bohnet}}, \bibinfo {author} {\bibfnamefont {K.~C.}\ \bibnamefont {Cox}},
  \bibinfo {author} {\bibfnamefont {M.~A.}\ \bibnamefont {Norcia}}, \bibinfo
  {author} {\bibfnamefont {J.~M.}\ \bibnamefont {Weiner}}, \bibinfo {author}
  {\bibfnamefont {Z.}~\bibnamefont {Chen}}, \ and\ \bibinfo {author}
  {\bibfnamefont {J.~K.}\ \bibnamefont {Thompson}},\ }\href@noop {} {\bibfield
  {journal} {\bibinfo  {journal} {Nature Photonics}\ }\textbf {\bibinfo
  {volume} {8}},\ \bibinfo {pages} {731} (\bibinfo {year} {2014})}\BibitemShut
  {NoStop}%
\bibitem [{\citenamefont {Tanaka}\ \emph {et~al.}(2015)\citenamefont {Tanaka},
  \citenamefont {Knott}, \citenamefont {Matsuzaki}, \citenamefont {Dooley},
  \citenamefont {Yamaguchi}, \citenamefont {Munro},\ and\ \citenamefont
  {Saito}}]{tanaka2015proposed}%
  \BibitemOpen
  \bibfield  {author} {\bibinfo {author} {\bibfnamefont {T.}~\bibnamefont
  {Tanaka}}, \bibinfo {author} {\bibfnamefont {P.}~\bibnamefont {Knott}},
  \bibinfo {author} {\bibfnamefont {Y.}~\bibnamefont {Matsuzaki}}, \bibinfo
  {author} {\bibfnamefont {S.}~\bibnamefont {Dooley}}, \bibinfo {author}
  {\bibfnamefont {H.}~\bibnamefont {Yamaguchi}}, \bibinfo {author}
  {\bibfnamefont {W.~J.}\ \bibnamefont {Munro}}, \ and\ \bibinfo {author}
  {\bibfnamefont {S.}~\bibnamefont {Saito}},\ }\href@noop {} {\bibfield
  {journal} {\bibinfo  {journal} {Physical review letters}\ }\textbf {\bibinfo
  {volume} {115}},\ \bibinfo {pages} {170801} (\bibinfo {year}
  {2015})}\BibitemShut {NoStop}%
\bibitem [{\citenamefont {Dooley}\ \emph {et~al.}(2016)\citenamefont {Dooley},
  \citenamefont {Yukawa}, \citenamefont {Matsuzaki}, \citenamefont {Knee},
  \citenamefont {Munro},\ and\ \citenamefont {Nemoto}}]{dooley2016hybrid}%
  \BibitemOpen
  \bibfield  {author} {\bibinfo {author} {\bibfnamefont {S.}~\bibnamefont
  {Dooley}}, \bibinfo {author} {\bibfnamefont {E.}~\bibnamefont {Yukawa}},
  \bibinfo {author} {\bibfnamefont {Y.}~\bibnamefont {Matsuzaki}}, \bibinfo
  {author} {\bibfnamefont {G.~C.}\ \bibnamefont {Knee}}, \bibinfo {author}
  {\bibfnamefont {W.~J.}\ \bibnamefont {Munro}}, \ and\ \bibinfo {author}
  {\bibfnamefont {K.}~\bibnamefont {Nemoto}},\ }\href@noop {} {\bibfield
  {journal} {\bibinfo  {journal} {New Journal of Physics}\ }\textbf {\bibinfo
  {volume} {18}},\ \bibinfo {pages} {053011} (\bibinfo {year}
  {2016})}\BibitemShut {NoStop}%
\bibitem [{\citenamefont {Davis}\ \emph {et~al.}(2016)\citenamefont {Davis},
  \citenamefont {Bentsen},\ and\ \citenamefont
  {Schleier-Smith}}]{davis2016approaching}%
  \BibitemOpen
  \bibfield  {author} {\bibinfo {author} {\bibfnamefont {E.}~\bibnamefont
  {Davis}}, \bibinfo {author} {\bibfnamefont {G.}~\bibnamefont {Bentsen}}, \
  and\ \bibinfo {author} {\bibfnamefont {M.}~\bibnamefont {Schleier-Smith}},\
  }\href@noop {} {\bibfield  {journal} {\bibinfo  {journal} {Physical review
  letters}\ }\textbf {\bibinfo {volume} {116}},\ \bibinfo {pages} {053601}
  (\bibinfo {year} {2016})}\BibitemShut {NoStop}%
\bibitem [{\citenamefont {Matsuzaki}\ \emph {et~al.}(2018)\citenamefont
  {Matsuzaki}, \citenamefont {Benjamin}, \citenamefont {Nakayama},
  \citenamefont {Saito},\ and\ \citenamefont {Munro}}]{matsuzaki2018quantum}%
  \BibitemOpen
  \bibfield  {author} {\bibinfo {author} {\bibfnamefont {Y.}~\bibnamefont
  {Matsuzaki}}, \bibinfo {author} {\bibfnamefont {S.}~\bibnamefont {Benjamin}},
  \bibinfo {author} {\bibfnamefont {S.}~\bibnamefont {Nakayama}}, \bibinfo
  {author} {\bibfnamefont {S.}~\bibnamefont {Saito}}, \ and\ \bibinfo {author}
  {\bibfnamefont {W.~J.}\ \bibnamefont {Munro}},\ }\href@noop {} {\bibfield
  {journal} {\bibinfo  {journal} {Physical review letters}\ }\textbf {\bibinfo
  {volume} {120}},\ \bibinfo {pages} {140501} (\bibinfo {year}
  {2018})}\BibitemShut {NoStop}%
\bibitem [{\citenamefont {Huber}\ \emph {et~al.}(2008)\citenamefont {Huber},
  \citenamefont {Koshnick}, \citenamefont {Bluhm}, \citenamefont {Archuleta},
  \citenamefont {Azua}, \citenamefont {Bj{\"o}rnsson}, \citenamefont {Gardner},
  \citenamefont {Halloran}, \citenamefont {Lucero},\ and\ \citenamefont
  {Moler}}]{huber2008gradiometric}%
  \BibitemOpen
  \bibfield  {author} {\bibinfo {author} {\bibfnamefont {M.~E.}\ \bibnamefont
  {Huber}}, \bibinfo {author} {\bibfnamefont {N.~C.}\ \bibnamefont {Koshnick}},
  \bibinfo {author} {\bibfnamefont {H.}~\bibnamefont {Bluhm}}, \bibinfo
  {author} {\bibfnamefont {L.~J.}\ \bibnamefont {Archuleta}}, \bibinfo {author}
  {\bibfnamefont {T.}~\bibnamefont {Azua}}, \bibinfo {author} {\bibfnamefont
  {P.~G.}\ \bibnamefont {Bj{\"o}rnsson}}, \bibinfo {author} {\bibfnamefont
  {B.~W.}\ \bibnamefont {Gardner}}, \bibinfo {author} {\bibfnamefont {S.~T.}\
  \bibnamefont {Halloran}}, \bibinfo {author} {\bibfnamefont {E.~A.}\
  \bibnamefont {Lucero}}, \ and\ \bibinfo {author} {\bibfnamefont {K.~A.}\
  \bibnamefont {Moler}},\ }\href@noop {} {\bibfield  {journal} {\bibinfo
  {journal} {Review of Scientific Instruments}\ }\textbf {\bibinfo {volume}
  {79}},\ \bibinfo {pages} {053704} (\bibinfo {year} {2008})}\BibitemShut
  {NoStop}%
\bibitem [{\citenamefont {Ramsden}(2011)}]{ramsden2011hall}%
  \BibitemOpen
  \bibfield  {author} {\bibinfo {author} {\bibfnamefont {E.}~\bibnamefont
  {Ramsden}},\ }\href@noop {} {\emph {\bibinfo {title} {Hall-effect Sensors:
  Theory and Application}}}\ (\bibinfo  {publisher} {Elsevier},\ \bibinfo
  {year} {2011})\BibitemShut {NoStop}%
\bibitem [{\citenamefont {Poggio}\ and\ \citenamefont
  {Degen}(2010)}]{poggio2010force}%
  \BibitemOpen
  \bibfield  {author} {\bibinfo {author} {\bibfnamefont {M.}~\bibnamefont
  {Poggio}}\ and\ \bibinfo {author} {\bibfnamefont {C.~L.}\ \bibnamefont
  {Degen}},\ }\href@noop {} {\bibfield  {journal} {\bibinfo  {journal}
  {Nanotechnology}\ }\textbf {\bibinfo {volume} {21}},\ \bibinfo {pages}
  {342001} (\bibinfo {year} {2010})}\BibitemShut {NoStop}%
\bibitem [{\citenamefont {Happer}\ and\ \citenamefont
  {Tang}(1973)}]{happer1973spin}%
  \BibitemOpen
  \bibfield  {author} {\bibinfo {author} {\bibfnamefont {W.}~\bibnamefont
  {Happer}}\ and\ \bibinfo {author} {\bibfnamefont {H.}~\bibnamefont {Tang}},\
  }\href@noop {} {\bibfield  {journal} {\bibinfo  {journal} {Physical Review
  Letters}\ }\textbf {\bibinfo {volume} {31}},\ \bibinfo {pages} {273}
  (\bibinfo {year} {1973})}\BibitemShut {NoStop}%
\bibitem [{\citenamefont {Allred}\ \emph {et~al.}(2002)\citenamefont {Allred},
  \citenamefont {Lyman}, \citenamefont {Kornack},\ and\ \citenamefont
  {Romalis}}]{allred2002high}%
  \BibitemOpen
  \bibfield  {author} {\bibinfo {author} {\bibfnamefont {J.~C.}\ \bibnamefont
  {Allred}}, \bibinfo {author} {\bibfnamefont {R.~N.}\ \bibnamefont {Lyman}},
  \bibinfo {author} {\bibfnamefont {T.~W.}\ \bibnamefont {Kornack}}, \ and\
  \bibinfo {author} {\bibfnamefont {M.~V.}\ \bibnamefont {Romalis}},\
  }\href@noop {} {\bibfield  {journal} {\bibinfo  {journal} {Physical Review
  Letters}\ }\textbf {\bibinfo {volume} {89}},\ \bibinfo {pages} {130801}
  (\bibinfo {year} {2002})}\BibitemShut {NoStop}%
\bibitem [{\citenamefont {Dang}\ \emph {et~al.}(2010)\citenamefont {Dang},
  \citenamefont {Maloof},\ and\ \citenamefont {Romalis}}]{dang2010ultrahigh}%
  \BibitemOpen
  \bibfield  {author} {\bibinfo {author} {\bibfnamefont {H.}~\bibnamefont
  {Dang}}, \bibinfo {author} {\bibfnamefont {A.}~\bibnamefont {Maloof}}, \ and\
  \bibinfo {author} {\bibfnamefont {M.}~\bibnamefont {Romalis}},\ }\href@noop
  {} {\bibfield  {journal} {\bibinfo  {journal} {Applied Physics Letters}\
  }\textbf {\bibinfo {volume} {97}},\ \bibinfo {pages} {151110} (\bibinfo
  {year} {2010})}\BibitemShut {NoStop}%
\bibitem [{\citenamefont {Bal}\ \emph {et~al.}(2012)\citenamefont {Bal},
  \citenamefont {Deng}, \citenamefont {Orgiazzi}, \citenamefont {Ong},\ and\
  \citenamefont {Lupascu}}]{bal2012ultrasensitive}%
  \BibitemOpen
  \bibfield  {author} {\bibinfo {author} {\bibfnamefont {M.}~\bibnamefont
  {Bal}}, \bibinfo {author} {\bibfnamefont {C.}~\bibnamefont {Deng}}, \bibinfo
  {author} {\bibfnamefont {J.-L.}\ \bibnamefont {Orgiazzi}}, \bibinfo {author}
  {\bibfnamefont {F.}~\bibnamefont {Ong}}, \ and\ \bibinfo {author}
  {\bibfnamefont {A.}~\bibnamefont {Lupascu}},\ }\href@noop {} {\bibfield
  {journal} {\bibinfo  {journal} {Nature communications}\ }\textbf {\bibinfo
  {volume} {3}},\ \bibinfo {pages} {1324} (\bibinfo {year} {2012})}\BibitemShut
  {NoStop}%
\bibitem [{\citenamefont {Toida}\ \emph {et~al.}(2017)\citenamefont {Toida},
  \citenamefont {Matsuzaki}, \citenamefont {Kakuyanagi}, \citenamefont {Zhu},
  \citenamefont {Munro}, \citenamefont {Yamaguchi},\ and\ \citenamefont
  {Saito}}]{toida2017electron}%
  \BibitemOpen
  \bibfield  {author} {\bibinfo {author} {\bibfnamefont {H.}~\bibnamefont
  {Toida}}, \bibinfo {author} {\bibfnamefont {Y.}~\bibnamefont {Matsuzaki}},
  \bibinfo {author} {\bibfnamefont {K.}~\bibnamefont {Kakuyanagi}}, \bibinfo
  {author} {\bibfnamefont {X.}~\bibnamefont {Zhu}}, \bibinfo {author}
  {\bibfnamefont {W.~J.}\ \bibnamefont {Munro}}, \bibinfo {author}
  {\bibfnamefont {H.}~\bibnamefont {Yamaguchi}}, \ and\ \bibinfo {author}
  {\bibfnamefont {S.}~\bibnamefont {Saito}},\ }\href@noop {} {\bibfield
  {journal} {\bibinfo  {journal} {arXiv preprint arXiv:1711.10148}\ } (\bibinfo
  {year} {2017})}\BibitemShut {NoStop}%
\bibitem [{\citenamefont {Acosta}\ \emph {et~al.}(2009)\citenamefont {Acosta},
  \citenamefont {Bauch}, \citenamefont {Ledbetter}, \citenamefont {Santori},
  \citenamefont {Fu}, \citenamefont {Barclay}, \citenamefont {Beausoleil},
  \citenamefont {Linget}, \citenamefont {Roch}, \citenamefont {Treussart} \emph
  {et~al.}}]{acosta2009diamonds}%
  \BibitemOpen
  \bibfield  {author} {\bibinfo {author} {\bibfnamefont {V.~M.}\ \bibnamefont
  {Acosta}}, \bibinfo {author} {\bibfnamefont {E.}~\bibnamefont {Bauch}},
  \bibinfo {author} {\bibfnamefont {M.~P.}\ \bibnamefont {Ledbetter}}, \bibinfo
  {author} {\bibfnamefont {C.}~\bibnamefont {Santori}}, \bibinfo {author}
  {\bibfnamefont {K.-M.~C.}\ \bibnamefont {Fu}}, \bibinfo {author}
  {\bibfnamefont {P.~E.}\ \bibnamefont {Barclay}}, \bibinfo {author}
  {\bibfnamefont {R.~G.}\ \bibnamefont {Beausoleil}}, \bibinfo {author}
  {\bibfnamefont {H.}~\bibnamefont {Linget}}, \bibinfo {author} {\bibfnamefont
  {J.~F.}\ \bibnamefont {Roch}}, \bibinfo {author} {\bibfnamefont
  {F.}~\bibnamefont {Treussart}},  \emph {et~al.},\ }\href@noop {} {\bibfield
  {journal} {\bibinfo  {journal} {Physical Review B}\ }\textbf {\bibinfo
  {volume} {80}},\ \bibinfo {pages} {115202} (\bibinfo {year}
  {2009})}\BibitemShut {NoStop}%
\bibitem [{\citenamefont {Balasubramanian}\ \emph {et~al.}(2009)\citenamefont
  {Balasubramanian}, \citenamefont {Neumann}, \citenamefont {Twitchen},
  \citenamefont {Markham}, \citenamefont {Kolesov}, \citenamefont {Mizuochi},
  \citenamefont {Isoya}, \citenamefont {Achard}, \citenamefont {Beck},
  \citenamefont {Tissler} \emph {et~al.}}]{balasubramanian2009ultralong}%
  \BibitemOpen
  \bibfield  {author} {\bibinfo {author} {\bibfnamefont {G.}~\bibnamefont
  {Balasubramanian}}, \bibinfo {author} {\bibfnamefont {P.}~\bibnamefont
  {Neumann}}, \bibinfo {author} {\bibfnamefont {D.}~\bibnamefont {Twitchen}},
  \bibinfo {author} {\bibfnamefont {M.}~\bibnamefont {Markham}}, \bibinfo
  {author} {\bibfnamefont {R.}~\bibnamefont {Kolesov}}, \bibinfo {author}
  {\bibfnamefont {N.}~\bibnamefont {Mizuochi}}, \bibinfo {author}
  {\bibfnamefont {J.}~\bibnamefont {Isoya}}, \bibinfo {author} {\bibfnamefont
  {J.}~\bibnamefont {Achard}}, \bibinfo {author} {\bibfnamefont
  {J.}~\bibnamefont {Beck}}, \bibinfo {author} {\bibfnamefont {J.}~\bibnamefont
  {Tissler}},  \emph {et~al.},\ }\href@noop {} {\bibfield  {journal} {\bibinfo
  {journal} {Nature materials}\ }\textbf {\bibinfo {volume} {8}},\ \bibinfo
  {pages} {383} (\bibinfo {year} {2009})}\BibitemShut {NoStop}%
\bibitem [{\citenamefont {Dolde}\ \emph {et~al.}(2011)\citenamefont {Dolde},
  \citenamefont {Fedder}, \citenamefont {Doherty}, \citenamefont {N{\"o}bauer},
  \citenamefont {Rempp}, \citenamefont {Balasubramanian}, \citenamefont {Wolf},
  \citenamefont {Reinhard}, \citenamefont {Hollenberg}, \citenamefont {Jelezko}
  \emph {et~al.}}]{nvcenter}%
  \BibitemOpen
  \bibfield  {author} {\bibinfo {author} {\bibfnamefont {F.}~\bibnamefont
  {Dolde}}, \bibinfo {author} {\bibfnamefont {H.}~\bibnamefont {Fedder}},
  \bibinfo {author} {\bibfnamefont {M.~W.}\ \bibnamefont {Doherty}}, \bibinfo
  {author} {\bibfnamefont {T.}~\bibnamefont {N{\"o}bauer}}, \bibinfo {author}
  {\bibfnamefont {F.}~\bibnamefont {Rempp}}, \bibinfo {author} {\bibfnamefont
  {G.}~\bibnamefont {Balasubramanian}}, \bibinfo {author} {\bibfnamefont
  {T.}~\bibnamefont {Wolf}}, \bibinfo {author} {\bibfnamefont {F.}~\bibnamefont
  {Reinhard}}, \bibinfo {author} {\bibfnamefont {L.~C.}\ \bibnamefont
  {Hollenberg}}, \bibinfo {author} {\bibfnamefont {F.}~\bibnamefont {Jelezko}},
   \emph {et~al.},\ }\href@noop {} {\bibfield  {journal} {\bibinfo  {journal}
  {Nature Physics}\ }\textbf {\bibinfo {volume} {7}},\ \bibinfo {pages} {459}
  (\bibinfo {year} {2011})}\BibitemShut {NoStop}%
\bibitem [{\citenamefont {Ishikawa}\ \emph {et~al.}(2012)\citenamefont
  {Ishikawa}, \citenamefont {Fu}, \citenamefont {Santori}, \citenamefont
  {Acosta}, \citenamefont {Beausoleil}, \citenamefont {Watanabe}, \citenamefont
  {Shikata},\ and\ \citenamefont {Itoh}}]{ishikawa2012optical}%
  \BibitemOpen
  \bibfield  {author} {\bibinfo {author} {\bibfnamefont {T.}~\bibnamefont
  {Ishikawa}}, \bibinfo {author} {\bibfnamefont {K.-M.~C.}\ \bibnamefont {Fu}},
  \bibinfo {author} {\bibfnamefont {C.}~\bibnamefont {Santori}}, \bibinfo
  {author} {\bibfnamefont {V.~M.}\ \bibnamefont {Acosta}}, \bibinfo {author}
  {\bibfnamefont {R.~G.}\ \bibnamefont {Beausoleil}}, \bibinfo {author}
  {\bibfnamefont {H.}~\bibnamefont {Watanabe}}, \bibinfo {author}
  {\bibfnamefont {S.}~\bibnamefont {Shikata}}, \ and\ \bibinfo {author}
  {\bibfnamefont {K.~M.}\ \bibnamefont {Itoh}},\ }\href@noop {} {\bibfield
  {journal} {\bibinfo  {journal} {Nano letters}\ }\textbf {\bibinfo {volume}
  {12}},\ \bibinfo {pages} {2083} (\bibinfo {year} {2012})}\BibitemShut
  {NoStop}%
\bibitem [{Note1()}]{Note1}%
  \BibitemOpen
  \bibinfo {note} {As done in the field of quantum metrology, we focus on the
  scaling, neglecting the constant factor.}\BibitemShut {Stop}%
\bibitem [{\citenamefont {Palma}\ \emph {et~al.}(1996)\citenamefont {Palma},
  \citenamefont {Suominen},\ and\ \citenamefont {Ekert}}]{palma1996quantum}%
  \BibitemOpen
  \bibfield  {author} {\bibinfo {author} {\bibfnamefont {G.~M.}\ \bibnamefont
  {Palma}}, \bibinfo {author} {\bibfnamefont {K.-A.}\ \bibnamefont {Suominen}},
  \ and\ \bibinfo {author} {\bibfnamefont {A.}~\bibnamefont {Ekert}},\
  }\href@noop {} {\bibfield  {journal} {\bibinfo  {journal} {Proc. R. Soc.
  Lond. A}\ }\textbf {\bibinfo {volume} {452}},\ \bibinfo {pages} {567}
  (\bibinfo {year} {1996})}\BibitemShut {NoStop}%
\bibitem [{\citenamefont {Smirne}\ \emph {et~al.}(2016)\citenamefont {Smirne},
  \citenamefont {Ko{\l}ody{\'n}ski}, \citenamefont {Huelga},\ and\
  \citenamefont {Demkowicz-Dobrza{\'n}ski}}]{smirne2016ultimate}%
  \BibitemOpen
  \bibfield  {author} {\bibinfo {author} {\bibfnamefont {A.}~\bibnamefont
  {Smirne}}, \bibinfo {author} {\bibfnamefont {J.}~\bibnamefont
  {Ko{\l}ody{\'n}ski}}, \bibinfo {author} {\bibfnamefont {S.~F.}\ \bibnamefont
  {Huelga}}, \ and\ \bibinfo {author} {\bibfnamefont {R.}~\bibnamefont
  {Demkowicz-Dobrza{\'n}ski}},\ }\href@noop {} {\bibfield  {journal} {\bibinfo
  {journal} {Physical review letters}\ }\textbf {\bibinfo {volume} {116}},\
  \bibinfo {pages} {120801} (\bibinfo {year} {2016})}\BibitemShut {NoStop}%
\bibitem [{\citenamefont {Macieszczak}(2015)}]{macieszczak2015zeno}%
  \BibitemOpen
  \bibfield  {author} {\bibinfo {author} {\bibfnamefont {K.}~\bibnamefont
  {Macieszczak}},\ }\href@noop {} {\bibfield  {journal} {\bibinfo  {journal}
  {Physical Review A}\ }\textbf {\bibinfo {volume} {92}},\ \bibinfo {pages}
  {010102(R)} (\bibinfo {year} {2015})}\BibitemShut {NoStop}%
\bibitem [{\citenamefont {Schr{\"o}dinger}(1935)}]{schroedinger}%
  \BibitemOpen
  \bibfield  {author} {\bibinfo {author} {\bibfnamefont {E.}~\bibnamefont
  {Schr{\"o}dinger}},\ }\href@noop {} {\bibfield  {journal} {\bibinfo
  {journal} {Naturwissenschaften}\ }\textbf {\bibinfo {volume} {23}},\ \bibinfo
  {pages} {823} (\bibinfo {year} {1935})}\BibitemShut {NoStop}%
\bibitem [{\citenamefont {Greenberger}\ \emph {et~al.}(1990)\citenamefont
  {Greenberger}, \citenamefont {Horne}, \citenamefont {Shimony},\ and\
  \citenamefont {Zeilinger}}]{greenberger1990bell}%
  \BibitemOpen
  \bibfield  {author} {\bibinfo {author} {\bibfnamefont {D.~M.}\ \bibnamefont
  {Greenberger}}, \bibinfo {author} {\bibfnamefont {M.~A.}\ \bibnamefont
  {Horne}}, \bibinfo {author} {\bibfnamefont {A.}~\bibnamefont {Shimony}}, \
  and\ \bibinfo {author} {\bibfnamefont {A.}~\bibnamefont {Zeilinger}},\
  }\href@noop {} {\bibfield  {journal} {\bibinfo  {journal} {American Journal
  of Physics}\ }\textbf {\bibinfo {volume} {58}},\ \bibinfo {pages} {1131}
  (\bibinfo {year} {1990})}\BibitemShut {NoStop}%
\bibitem [{\citenamefont {Monz}\ \emph {et~al.}(2011)\citenamefont {Monz},
  \citenamefont {Schindler}, \citenamefont {Barreiro}, \citenamefont {Chwalla},
  \citenamefont {Nigg}, \citenamefont {Coish}, \citenamefont {Harlander},
  \citenamefont {H{\"a}nsel}, \citenamefont {Hennrich},\ and\ \citenamefont
  {Blatt}}]{monz201114}%
  \BibitemOpen
  \bibfield  {author} {\bibinfo {author} {\bibfnamefont {T.}~\bibnamefont
  {Monz}}, \bibinfo {author} {\bibfnamefont {P.}~\bibnamefont {Schindler}},
  \bibinfo {author} {\bibfnamefont {J.~T.}\ \bibnamefont {Barreiro}}, \bibinfo
  {author} {\bibfnamefont {M.}~\bibnamefont {Chwalla}}, \bibinfo {author}
  {\bibfnamefont {D.}~\bibnamefont {Nigg}}, \bibinfo {author} {\bibfnamefont
  {W.~A.}\ \bibnamefont {Coish}}, \bibinfo {author} {\bibfnamefont
  {M.}~\bibnamefont {Harlander}}, \bibinfo {author} {\bibfnamefont
  {W.}~\bibnamefont {H{\"a}nsel}}, \bibinfo {author} {\bibfnamefont
  {M.}~\bibnamefont {Hennrich}}, \ and\ \bibinfo {author} {\bibfnamefont
  {R.}~\bibnamefont {Blatt}},\ }\href@noop {} {\bibfield  {journal} {\bibinfo
  {journal} {Physical Review Letters}\ }\textbf {\bibinfo {volume} {106}},\
  \bibinfo {pages} {130506} (\bibinfo {year} {2011})}\BibitemShut {NoStop}%
\bibitem [{\citenamefont {DiCarlo}\ \emph {et~al.}(2010)\citenamefont
  {DiCarlo}, \citenamefont {Reed}, \citenamefont {Sun}, \citenamefont
  {Johnson}, \citenamefont {Chow}, \citenamefont {Gambetta}, \citenamefont
  {Frunzio}, \citenamefont {Girvin}, \citenamefont {Devoret},\ and\
  \citenamefont {Schoelkopf}}]{dicarlo2010preparation}%
  \BibitemOpen
  \bibfield  {author} {\bibinfo {author} {\bibfnamefont {L.}~\bibnamefont
  {DiCarlo}}, \bibinfo {author} {\bibfnamefont {M.~D.}\ \bibnamefont {Reed}},
  \bibinfo {author} {\bibfnamefont {L.}~\bibnamefont {Sun}}, \bibinfo {author}
  {\bibfnamefont {B.~R.}\ \bibnamefont {Johnson}}, \bibinfo {author}
  {\bibfnamefont {J.~M.}\ \bibnamefont {Chow}}, \bibinfo {author}
  {\bibfnamefont {J.~M.}\ \bibnamefont {Gambetta}}, \bibinfo {author}
  {\bibfnamefont {L.}~\bibnamefont {Frunzio}}, \bibinfo {author} {\bibfnamefont
  {S.~M.}\ \bibnamefont {Girvin}}, \bibinfo {author} {\bibfnamefont {M.~H.}\
  \bibnamefont {Devoret}}, \ and\ \bibinfo {author} {\bibfnamefont {R.~J.}\
  \bibnamefont {Schoelkopf}},\ }\href@noop {} {\bibfield  {journal} {\bibinfo
  {journal} {Nature}\ }\textbf {\bibinfo {volume} {467}},\ \bibinfo {pages}
  {574} (\bibinfo {year} {2010})}\BibitemShut {NoStop}%
\bibitem [{\citenamefont {Fr{\"o}wis}\ \emph {et~al.}(2018)\citenamefont
  {Fr{\"o}wis}, \citenamefont {Sekatski}, \citenamefont {D{\"u}r},
  \citenamefont {Gisin},\ and\ \citenamefont
  {Sangouard}}]{frowis2018macroscopic}%
  \BibitemOpen
  \bibfield  {author} {\bibinfo {author} {\bibfnamefont {F.}~\bibnamefont
  {Fr{\"o}wis}}, \bibinfo {author} {\bibfnamefont {P.}~\bibnamefont
  {Sekatski}}, \bibinfo {author} {\bibfnamefont {W.}~\bibnamefont {D{\"u}r}},
  \bibinfo {author} {\bibfnamefont {N.}~\bibnamefont {Gisin}}, \ and\ \bibinfo
  {author} {\bibfnamefont {N.}~\bibnamefont {Sangouard}},\ }\href@noop {}
  {\bibfield  {journal} {\bibinfo  {journal} {Reviews of Modern Physics}\
  }\textbf {\bibinfo {volume} {90}},\ \bibinfo {pages} {025004} (\bibinfo
  {year} {2018})}\BibitemShut {NoStop}%
\bibitem [{\citenamefont {Shimizu}\ and\ \citenamefont {Morimae}(2005)}]{q}%
  \BibitemOpen
  \bibfield  {author} {\bibinfo {author} {\bibfnamefont {A.}~\bibnamefont
  {Shimizu}}\ and\ \bibinfo {author} {\bibfnamefont {T.}~\bibnamefont
  {Morimae}},\ }\href@noop {} {\bibfield  {journal} {\bibinfo  {journal}
  {Physical Review Letters}\ }\textbf {\bibinfo {volume} {95}},\ \bibinfo
  {pages} {090401} (\bibinfo {year} {2005})}\BibitemShut {NoStop}%
\bibitem [{\citenamefont {Kitagawa}\ and\ \citenamefont
  {Ueda}(1993)}]{kitagawa1993squeezed}%
  \BibitemOpen
  \bibfield  {author} {\bibinfo {author} {\bibfnamefont {M.}~\bibnamefont
  {Kitagawa}}\ and\ \bibinfo {author} {\bibfnamefont {M.}~\bibnamefont
  {Ueda}},\ }\href@noop {} {\bibfield  {journal} {\bibinfo  {journal} {Physical
  Review A}\ }\textbf {\bibinfo {volume} {47}},\ \bibinfo {pages} {5138}
  (\bibinfo {year} {1993})}\BibitemShut {NoStop}%
\bibitem [{\citenamefont {Tatsuta}\ and\ \citenamefont
  {Shimizu}(2018)}]{tatsuta2018conversion}%
  \BibitemOpen
  \bibfield  {author} {\bibinfo {author} {\bibfnamefont {M.}~\bibnamefont
  {Tatsuta}}\ and\ \bibinfo {author} {\bibfnamefont {A.}~\bibnamefont
  {Shimizu}},\ }\href@noop {} {\bibfield  {journal} {\bibinfo  {journal}
  {Physical Review A}\ }\textbf {\bibinfo {volume} {97}},\ \bibinfo {pages}
  {012124} (\bibinfo {year} {2018})}\BibitemShut {NoStop}%
\bibitem [{\citenamefont {Fr{\"o}wis}\ and\ \citenamefont
  {D{\"u}r}(2012)}]{frowis2012measures}%
  \BibitemOpen
  \bibfield  {author} {\bibinfo {author} {\bibfnamefont {F.}~\bibnamefont
  {Fr{\"o}wis}}\ and\ \bibinfo {author} {\bibfnamefont {W.}~\bibnamefont
  {D{\"u}r}},\ }\href@noop {} {\bibfield  {journal} {\bibinfo  {journal} {New
  Journal of Physics}\ }\textbf {\bibinfo {volume} {14}},\ \bibinfo {pages}
  {093039} (\bibinfo {year} {2012})}\BibitemShut {NoStop}%
\bibitem [{\citenamefont {Morimae}(2010)}]{morimae2010superposition}%
  \BibitemOpen
  \bibfield  {author} {\bibinfo {author} {\bibfnamefont {T.}~\bibnamefont
  {Morimae}},\ }\href@noop {} {\bibfield  {journal} {\bibinfo  {journal}
  {Physical Review A}\ }\textbf {\bibinfo {volume} {81}},\ \bibinfo {pages}
  {010101(R)} (\bibinfo {year} {2010})}\BibitemShut {NoStop}%
\bibitem [{\citenamefont {Jeong}\ \emph {et~al.}(2015)\citenamefont {Jeong},
  \citenamefont {Kang},\ and\ \citenamefont
  {Kwon}}]{jeong2015characterizations}%
  \BibitemOpen
  \bibfield  {author} {\bibinfo {author} {\bibfnamefont {H.}~\bibnamefont
  {Jeong}}, \bibinfo {author} {\bibfnamefont {M.}~\bibnamefont {Kang}}, \ and\
  \bibinfo {author} {\bibfnamefont {H.}~\bibnamefont {Kwon}},\ }\href@noop {}
  {\bibfield  {journal} {\bibinfo  {journal} {Optics Communications}\ }\textbf
  {\bibinfo {volume} {337}},\ \bibinfo {pages} {12} (\bibinfo {year}
  {2015})}\BibitemShut {NoStop}%
\bibitem [{\citenamefont {Fr{\"o}wis}\ \emph {et~al.}(2015)\citenamefont
  {Fr{\"o}wis}, \citenamefont {Sangouard},\ and\ \citenamefont
  {Gisin}}]{frowis2015linking}%
  \BibitemOpen
  \bibfield  {author} {\bibinfo {author} {\bibfnamefont {F.}~\bibnamefont
  {Fr{\"o}wis}}, \bibinfo {author} {\bibfnamefont {N.}~\bibnamefont
  {Sangouard}}, \ and\ \bibinfo {author} {\bibfnamefont {N.}~\bibnamefont
  {Gisin}},\ }\href@noop {} {\bibfield  {journal} {\bibinfo  {journal} {Optics
  communications}\ }\textbf {\bibinfo {volume} {337}},\ \bibinfo {pages} {2}
  (\bibinfo {year} {2015})}\BibitemShut {NoStop}%
\bibitem [{\citenamefont {Abad}\ and\ \citenamefont
  {Karimipour}(2016)}]{PhysRevB.93.195127}%
  \BibitemOpen
  \bibfield  {author} {\bibinfo {author} {\bibfnamefont {T.}~\bibnamefont
  {Abad}}\ and\ \bibinfo {author} {\bibfnamefont {V.}~\bibnamefont
  {Karimipour}},\ }\href {\doibase 10.1103/PhysRevB.93.195127} {\bibfield
  {journal} {\bibinfo  {journal} {Phys. Rev. B}\ }\textbf {\bibinfo {volume}
  {93}},\ \bibinfo {pages} {195127} (\bibinfo {year} {2016})}\BibitemShut
  {NoStop}%
\bibitem [{\citenamefont {Park}\ \emph {et~al.}(2016)\citenamefont {Park},
  \citenamefont {Kang}, \citenamefont {Lee}, \citenamefont {Bang},
  \citenamefont {Lee},\ and\ \citenamefont {Jeong}}]{park2016quantum}%
  \BibitemOpen
  \bibfield  {author} {\bibinfo {author} {\bibfnamefont {C.-Y.}\ \bibnamefont
  {Park}}, \bibinfo {author} {\bibfnamefont {M.}~\bibnamefont {Kang}}, \bibinfo
  {author} {\bibfnamefont {C.-W.}\ \bibnamefont {Lee}}, \bibinfo {author}
  {\bibfnamefont {J.}~\bibnamefont {Bang}}, \bibinfo {author} {\bibfnamefont
  {S.-W.}\ \bibnamefont {Lee}}, \ and\ \bibinfo {author} {\bibfnamefont
  {H.}~\bibnamefont {Jeong}},\ }\href@noop {} {\bibfield  {journal} {\bibinfo
  {journal} {Physical Review A}\ }\textbf {\bibinfo {volume} {94}},\ \bibinfo
  {pages} {052105} (\bibinfo {year} {2016})}\BibitemShut {NoStop}%
\bibitem [{\citenamefont {Shimizu}\ \emph {et~al.}(2013)\citenamefont
  {Shimizu}, \citenamefont {Matsuzaki},\ and\ \citenamefont {Ukena}}]{speedup}%
  \BibitemOpen
  \bibfield  {author} {\bibinfo {author} {\bibfnamefont {A.}~\bibnamefont
  {Shimizu}}, \bibinfo {author} {\bibfnamefont {Y.}~\bibnamefont {Matsuzaki}},
  \ and\ \bibinfo {author} {\bibfnamefont {A.}~\bibnamefont {Ukena}},\
  }\href@noop {} {\bibfield  {journal} {\bibinfo  {journal} {Journal of the
  Physical Society of Japan}\ }\textbf {\bibinfo {volume} {82}},\ \bibinfo
  {pages} {054801} (\bibinfo {year} {2013})}\BibitemShut {NoStop}%
\bibitem [{\citenamefont {Shimizu}\ and\ \citenamefont
  {Miyadera}(2002)}]{shmzmiyadera2002}%
  \BibitemOpen
  \bibfield  {author} {\bibinfo {author} {\bibfnamefont {A.}~\bibnamefont
  {Shimizu}}\ and\ \bibinfo {author} {\bibfnamefont {T.}~\bibnamefont
  {Miyadera}},\ }\href {\doibase 10.1103/PhysRevLett.89.270403} {\bibfield
  {journal} {\bibinfo  {journal} {Phys. Rev. Lett.}\ }\textbf {\bibinfo
  {volume} {89}},\ \bibinfo {pages} {270403} (\bibinfo {year}
  {2002})}\BibitemShut {NoStop}%
\bibitem [{\citenamefont {Escher}\ \emph {et~al.}(2011)\citenamefont {Escher},
  \citenamefont {de~Matos~Filho},\ and\ \citenamefont
  {Davidovich}}]{escher2011general}%
  \BibitemOpen
  \bibfield  {author} {\bibinfo {author} {\bibfnamefont {B.}~\bibnamefont
  {Escher}}, \bibinfo {author} {\bibfnamefont {R.}~\bibnamefont
  {de~Matos~Filho}}, \ and\ \bibinfo {author} {\bibfnamefont {L.}~\bibnamefont
  {Davidovich}},\ }\href@noop {} {\bibfield  {journal} {\bibinfo  {journal}
  {Nature Physics}\ }\textbf {\bibinfo {volume} {7}},\ \bibinfo {pages} {406}
  (\bibinfo {year} {2011})}\BibitemShut {NoStop}%
\bibitem [{\citenamefont {Ko{\l}ody{\'n}ski}\ and\ \citenamefont
  {Demkowicz-Dobrza{\'n}ski}(2013)}]{kolodynski2013efficient}%
  \BibitemOpen
  \bibfield  {author} {\bibinfo {author} {\bibfnamefont {J.}~\bibnamefont
  {Ko{\l}ody{\'n}ski}}\ and\ \bibinfo {author} {\bibfnamefont {R.}~\bibnamefont
  {Demkowicz-Dobrza{\'n}ski}},\ }\href@noop {} {\bibfield  {journal} {\bibinfo
  {journal} {New Journal of Physics}\ }\textbf {\bibinfo {volume} {15}},\
  \bibinfo {pages} {073043} (\bibinfo {year} {2013})}\BibitemShut {NoStop}%
\bibitem [{\citenamefont {Hornberger}(2009)}]{hornberger2009introduction}%
  \BibitemOpen
  \bibfield  {author} {\bibinfo {author} {\bibfnamefont {K.}~\bibnamefont
  {Hornberger}},\ }in\ \href@noop {} {\emph {\bibinfo {booktitle} {Entanglement
  and Decoherence}}}\ (\bibinfo  {publisher} {Springer},\ \bibinfo {year}
  {2009})\ pp.\ \bibinfo {pages} {221--276}\BibitemShut {NoStop}%
\bibitem [{\citenamefont {Tyryshkin}\ \emph {et~al.}(2012)\citenamefont
  {Tyryshkin}, \citenamefont {Tojo}, \citenamefont {Morton}, \citenamefont
  {Riemann}, \citenamefont {Abrosimov}, \citenamefont {Becker}, \citenamefont
  {Pohl}, \citenamefont {Schenkel}, \citenamefont {Thewalt}, \citenamefont
  {Itoh},\ and\ \citenamefont {Lyon}}]{eec6e647c2824bbab76a44bb03dd0eeb}%
  \BibitemOpen
  \bibfield  {author} {\bibinfo {author} {\bibfnamefont {A.}~\bibnamefont
  {Tyryshkin}}, \bibinfo {author} {\bibfnamefont {S.}~\bibnamefont {Tojo}},
  \bibinfo {author} {\bibfnamefont {J.}~\bibnamefont {Morton}}, \bibinfo
  {author} {\bibfnamefont {H.}~\bibnamefont {Riemann}}, \bibinfo {author}
  {\bibfnamefont {N.}~\bibnamefont {Abrosimov}}, \bibinfo {author}
  {\bibfnamefont {P.}~\bibnamefont {Becker}}, \bibinfo {author} {\bibfnamefont
  {H.}~\bibnamefont {Pohl}}, \bibinfo {author} {\bibfnamefont {T.}~\bibnamefont
  {Schenkel}}, \bibinfo {author} {\bibfnamefont {M.}~\bibnamefont {Thewalt}},
  \bibinfo {author} {\bibfnamefont {K.}~\bibnamefont {Itoh}}, \ and\ \bibinfo
  {author} {\bibfnamefont {S.}~\bibnamefont {Lyon}},\ }\href {\doibase
  10.1038/nmat3182} {\bibfield  {journal} {\bibinfo  {journal} {Nature
  Materials}\ }\textbf {\bibinfo {volume} {11}},\ \bibinfo {pages} {143}
  (\bibinfo {year} {2012})}\BibitemShut {NoStop}%
\bibitem [{\citenamefont {Baudenbacher}\ \emph {et~al.}(2003)\citenamefont
  {Baudenbacher}, \citenamefont {Fong}, \citenamefont {Holzer},\ and\
  \citenamefont {Radparvar}}]{baudenbacher2003monolithic}%
  \BibitemOpen
  \bibfield  {author} {\bibinfo {author} {\bibfnamefont {F.}~\bibnamefont
  {Baudenbacher}}, \bibinfo {author} {\bibfnamefont {L.}~\bibnamefont {Fong}},
  \bibinfo {author} {\bibfnamefont {J.}~\bibnamefont {Holzer}}, \ and\ \bibinfo
  {author} {\bibfnamefont {M.}~\bibnamefont {Radparvar}},\ }\href@noop {}
  {\bibfield  {journal} {\bibinfo  {journal} {Applied Physics Letters}\
  }\textbf {\bibinfo {volume} {82}},\ \bibinfo {pages} {3487} (\bibinfo {year}
  {2003})}\BibitemShut {NoStop}%
\end{thebibliography}%

\end{document}